\documentclass[preprint,notoc]{JHEP3} 
\usepackage{epsfig,multicol}
\usepackage{amssymb}
\usepackage{graphics}

\usepackage{amsmath,latexsym}
\def\beq{\begin{eqnarray}}    
\def\eeq{\end{eqnarray}}      

\newcommand{\OM}{\Omega_M}
\newcommand{\Om}{\Omega_m}
\newcommand{\Omo}{\Omega_m^0}

\newcommand{\OMo}{\Omega_{M}^0}
\newcommand{\ORo}{\Omega_{R}^0}
\newcommand{\OL}{\Omega_{\Lambda}}
\newcommand{\Oc}{\Omega_{c}}
\newcommand{\Oco}{\Omega_{c}^0}

\newcommand{\OLo}{\Omega_{\Lambda}^0}
\newcommand{\OX}{\Omega_{X}}
\newcommand{\OXo}{\Omega_{X}^0}

\newcommand{\OD}{\Omega_{D}}
\newcommand{\ODo}{\Omega_{D}^0}

\newcommand{\OK}{\Omega_K}
\newcommand{\OKo}{\Omega_{K}^0}

\newcommand{\rc}{\rho_c}
\newcommand{\rco}{\rho_{c}^0}

\newcommand{\rM}{\rho_M}
\newcommand{\rmr}{\rho_m}
\newcommand{\rMo}{\rho_{M}^0}

\newcommand{\rR}{\rho_R}
\newcommand{\rD}{\rho_D}
\newcommand{\rDo}{\rho_{D}^0}
\newcommand{\rX}{\rho_X}
\newcommand{\pX}{p_X}
\newcommand{\wX}{\omega_X}
\newcommand{\wm}{\omega_m}
\newcommand{\wR}{\omega_R}
\newcommand{\aR}{\alpha_R}
\newcommand{\amr}{\alpha_m}
\newcommand{\aef}{\alpha_e}
\newcommand{\aX}{\alpha_X}
\newcommand{\rL}{\rho_{\CC}}
\newcommand{\pL}{p_{\CC}}
\newcommand{\rLo}{\rho_{\CC}^0}
\newcommand{\pD}{p_D}
\newcommand{\wD}{\omega_D}
\newcommand{\zm}{z_{\rm max}}

\newcommand{\CC}{\Lambda}
\newcommand{\CCo}{\Lambda_0}

\newcommand{\we}{\omega_{e}}
\newcommand{\re}{r_{\epsilon}}

\newcommand{\lu}{\lambda_1}
\newcommand{\ld}{\lambda_2}
\newcommand{\lt}{\lambda_3}

\title{$\CC$XCDM: a cosmon model solution to
the cosmological coincidence problem?}
\author{\hspace{0.3cm}
Javier Grande $^{1}$, Joan Sol\`a $^{1,2}$, Hrvoje
\v{S}tefan\v{c}i\'{c} $^{1}$\,\thanks{On leave of absence from
the Theoretical Physics Division, Rudjer Bo\v{s}kovi\'{c}
Institute, Zagreb, Croatia.}\\
$^{1}\,$ High Energy Physics Group, Dep. ECM,  Univ. de
Barcelona,\\ \hspace{0.2cm}
Diagonal 647, 08028 Barcelona, Catalonia, Spain\\
\hspace{0.2cm} E-mails: jgrande@ecm.ub.es, sola@ifae.es, stefancic@ecm.ub.es\\
$^{2}\,$ C.E.R. for Astrophysics, Particle Physics and
Cosmology\,\thanks{Associated with Instituto de Ciencias del
Espacio-CSIC.}}

\preprint{ UB-ECM-PF-06/12\\ IRB-TH-1/06}

\abstract{We consider the possibility that the total dark energy
(DE) of the Universe is made out of two dynamical components of
different nature: a variable cosmological term, $\CC$, and a
dynamical ``cosmon'', $X$, possibly interacting with $\CC$ but
not with matter -- which remains conserved. We call this scenario
the $\CC$XCDM model. One possibility for $X$ would be a scalar
field $\chi$, but it is not the only one. The overall equation of
state (EOS) of the $\CC$XCDM model can effectively appear as
quintessence or phantom energy depending on the mixture of the two
components. Both the dynamics of $\CC$ and of $X$ could be linked
to high energy effects near the Planck scale. In the case of $\CC$
it may be related to the running of this parameter under quantum
effects, whereas $X$ might be identified with some fundamental
field (say, a dilaton)  left over as a low-energy ``relic'' by
e.g. string theory. We find that the dynamics of the $\CC$XCDM
model can trigger a future stopping of the Universe expansion and
can keep the ratio $\rD/\rmr$  (DE density to matter-radiation
density) bounded and of order $1$. Therefore, the model could
explain the so-called ``cosmological coincidence problem''. This
is in part related to the possibility that the present value of
the cosmological term can be $\CC_0<0$ in this framework (the
current \textit{total} DE density nevertheless being positive).
However, a cosmic halt could occur even if $\CC_0>0$ because of
the peculiar behavior of $X$ as ``Phantom Matter''. We describe
various cosmological scenarios made possible by the composite and
dynamical nature of $\CC$XCDM, and discuss in detail their impact
on the cosmological coincidence problem.}


\begin{document}

\section{Introduction}
\label{sect:Intro}

Modern experimental cosmology offers us a strikingly accurate
picture of our Universe which has been dubbed the
\textit{cosmological concordance model}, or standard $\CC$CDM
model\,\cite{Peebles84}. It is characterized by an essentially
zero value of the spatial curvature parameter and a
non-vanishing, but positive, value of the cosmological term (CT),
$\CC$, in Einstein's equations. The $\CC$CDM model supersedes the
old Einstein-de Sitter (or critical-density) model, which has
remained as the standard cold dark matter (CDM) cosmological model
for many years. The CDM model has also zero curvature but at the
same time it has zero CT ($\CC=0$), in contradiction with present
observations. Evidence for the values of the cosmological
parameters within the new concordance model comes both from
tracing the rate of expansion of the Universe with high-z Type Ia
supernovae experiments and from the precise measurement of the
anisotropies in the cosmic microwave background (CMB)
radiation\,\cite{Supernovae,WMAP03,LSS}. These experiments seem
to agree that $70\%\ (\OLo)$ of the total energy density of the
Universe at the present time should be encoded in the form of
vacuum energy density $\rL$ or, in general, as a new form of
energy density, $\rD$, which has been dubbed ``dark energy'' (DE)
because of our ignorance about its ultimate
nature\,\cite{TurnerWhite}. The remaining $30\%\
(\OMo\equiv\Omega_{\rm CDM}^0+\Omega_{\rm B}^0)$ of the present
energy budget is essentially made out of cold dark matter, with
only a scanty $\Omega_{\rm B}^0\lesssim 5\%$ consisting of
baryons.

At present the $\CC$CDM model is perfectly consistent with all
existing data\,\cite{Supernovae,WMAP03,LSS}, including the lately
released three year WMAP results\,\cite{WMAP3Y}. In particular,
this means that a non-vanishing and constant value $\CC=\CCo$ does
fit the data reasonably well\,\cite{Linde}. However, when
analyzing the pressure and density $(\pD,\rD)$ of the mysterious
DE fluid (presumably responsible for the accelerated expansion of
the Universe), some recent studies\,\cite{Alam,Hannestad,Jassal1}
seem to suggest that its equation-of-state (EOS) parameter,
$\wD\equiv\pD/\rho_D$, could be evolving with time or
(cosmological) redshift, $\wD=\wD(z)$, showing a significant
departure from the naive expectation $\wD=-1$ associated to the
CT. Although this potential variation seems to be dependent on
the kind of data set used, at present it cannot be ruled
out\,\cite{Jassal2,WMAP3Y}. In some cases it could even
accommodate $\wD<-1$, namely a ``phantom-like behavior''(see
below)\,\cite{Phantom}. In the near future we expect to get more
accurate measurements of the dark energy EOS from various
independent sources, including galaxy clusters, weak gravitational
lensing, supernova distances and CMB anisotropies at small angular
resolution. The main experiments that should provide this
information are SNAP\,\cite{SNAP}, PLANCK\,\cite{PLANCK} and
DES\,\cite{DES}. If the overall result from these experiments
would be that the DE density is constant (i.e. not evolving at all
with the cosmological redshift), then $\rD$ could just be
identified with the constant energy density $\rLo\sim
10^{-47}\,GeV^4$ associated to the cosmological constant
$\CC_0=8\,\pi\,G\,\rLo$ of the standard $\CC$CDM model. Even
though this could be looked upon as the simplest and most
economical explanation for the DE, it should not be regarded as
more satisfactory, for it would not shed any light on the physical
interpretation of $\CC$ in Einstein's equations. In particular,
the notion of $\CC$ as being a constant vacuum energy density
throughout the entire history of the Universe clashes violently
with all known predictions from quantum field theory (QFT),
including our cherished Standard Model of strong and electroweak
interactions. This leads to the famous cosmological constant
problem (CCP), the biggest conundrum in Theoretical Physics ever
\,\cite{weinRMP,CCRev,DEPaddy,Copeland06}\footnote{See also Ref.
\cite{JHEPCC1,SantFeliu} for a summarized presentation of the
CCP.}. While we do not have at present a clear idea of how to
associate the value of the vacuum energy to the CT, a dynamical
picture for the DE looks much more promising. For, if $\rD=\rD(t)$
is a function of the cosmic time, or equivalently of the
cosmological redshift $\rD=\rD(z)$, then there is a hope that we
can explain why it has the particular value $\rDo\sim
10^{-47}\,GeV^4$ we have measured today and why its value may have
been very different from the one (possibly much larger) that
$\rD$ took at early times, and the (tinier?) value it may take in
the remote future (including a potential change of sign). This
possibility has been suggested in different contexts
\,\cite{Xinesos,RGTypeIa} as a means for alleviating certain
problems with the formulation of asymptotic states in string
theory with positive $\CC$ \,\cite{Hellerman}.

The notion of a variable vacuum energy is rather old and it was
originally implemented in terms of dynamical scalar fields
\,\cite{Dolgov}. For example, the ``cosmon'' field introduced in
\cite{PSW} aimed at a Peccei-Quinn-like adjustment mechanism based
on a dynamical selection of the vacuum state at zero VEV of the
potential, $<V>=0$. More recently these ideas have been exploited
profusely in various forms, such as the so-called ``quintessence''
scalar fields and the like\,\cite{Ratra,quintessence}, ``phantom''
fields\,\cite{Phantom}, braneworld models\,\cite{braneworld},
Chaplygin gas\,\cite{Chaplygin}, and many other recent ideas like
holographic dark energy, cosmic strings, domain walls etc (see
e.g.  \,\cite{CCRev,DEPaddy,Copeland06} and references therein),
including some recently resurrected old ideas on adjusting
mechanisms\,\cite{Barr}, and also possible connections between the
DE and neutrino physics\,\cite{cosm,Peccei} or the existence of
extremely light quanta\,\cite{cosm,Bohmer}. If the equation of
state $\pD=\wD\,\rD$ describes some scalar field $\chi$ (perhaps
the most paradigmatic scenario for the dynamical DE models), and
the EOS parameter $\wD$ lies in the interval $-1<\wD<-1/3$, then
the field $\chi$ corresponds to standard
quintessence\,\cite{quintessence}; if, however, $\wD<-1$ then
$\chi$ is called a ``phantom field''\,\cite{Phantom} because this
possibility is non-canonical in QFT (namely it enforces the
coefficient in its kinetic energy term to be negative) and
violates the weak and (\textit{a fortiori}) the dominant energy
conditions\,\cite{HawkingEOS}.

Let us emphasize that the EOS parameter $\wD$ could be an
effective one\,\cite{LinderEff,SS1,SS2}, and hence we had better
call it $\we$. Actually, the effective $\we$ could result from a
model which does not even contain a single scalar field as the
direct physical support of the DE. For example, in Ref.\cite{SS1}
it has been shown that a running $\CC$ model leads to an
effective EOS, $\pD=\we\,\rD$, which may behave as quintessence
or even as a phantom-like fluid. Remarkably this feature has been
elevated to the category of a general theorem\,\cite{SS2}, which
states the following: any model based on Einstein's equations with
a variable $\CC=\CC(t)$ and/or $G=G(t)$ leads to an effective EOS
parameter $\we$ which can emulate the dynamical behavior of a
scalar field both in ``quintessence phase'' ($-1<\we<-1/3$) or in
``phantom phase'' ($\we<-1$) -- therefore always entailing a
crossing of the cosmological constant divide
$\we=-1$\,\cite{SS2}. For another example of this general
theorem, in a complementary situation where $G$ is variable but
$\rL$ is constant, see \cite{Guberina06}.

In spite of the virtues of cosmologies based on variable
cosmological parameters, it is convenient to study the possibility
of having a composite DE model involving additional ingredients.
As we shall see, this may help to smooth out other acute
cosmological problems. For instance, consider the so-called
``cosmological coincidence problem''\,\cite{Steinhardt}, to wit:
why do we find ourselves in an epoch $t=t_0$ where the DE density
is similar to the matter density ($\rD(t_0)\simeq\rM(t_0)$)?  In
the $\CC$CDM model this is an especially troublesome problem
because $\rL$ remains constant throughout the entire history of
the Universe. In the ordinary dynamical DE models the problem has
also its own difficulties. Thus, in a typical quintessence model
the matter-radiation energy density $\rmr=\rM+\rR$ decreases (with
the expansion) faster than the DE density and we expect that in
the early epochs $\rmr\gg\rD$, whereas at present and in the
future $\rmr\ll\rD$. Therefore, why do we just happen to live in
an epoch $t_0$ where the two functions $\rD(t_0)\simeq\rmr(t_0)$?
Is this a mere coincidence or there is some other, more
convincing, reason?  The next question of course is: can we
devise a model where the ratio $\rmr/\rD$ stays bounded (perhaps
even not too far from $1$) in essentially the entire span of the
Universe lifetime? This is certainly \textit{not} possible
neither in the standard $\CC$CDM model, nor in standard
quintessence models\,\footnote{There is, however, the possibility
to use more complicated models with
interactive\,\cite{InteractingQ} or
oscillating\,\cite{OscillatingQ} quintessence (see also
\cite{Copeland06}), or to consider the probability that
$\rD/\rmr\sim 1$ in a phantom Universe\,\cite{Scherrer}.}. And it
is also impossible for a model whose DE consists only of a running
$\CC$\,\cite{RGTypeIa}. However, in this paper we will show that
we can produce a dynamical DE model with such a property.
Specifically, we investigate a minimal realization of a composite
DE model made out of just two components: a running $\CC$ and
another entity, $X$, which interacts with $\CC$. In this
framework matter and radiation will be canonically conserved, and
the running of $\CC$ is (as any parameter in QFT) tied to the
renormalization group (RG) in curved
space-time\,\cite{JHEPCC1,Book}. To illustrate this possibility
we adapt a type of cosmological RG model with running $\CC$ that
has been thoroughly studied in the
literature\,\cite{RGTypeIa}\,\footnote{For the recent literature
on RG models supporting the idea of running cosmological
parameters, whether in QFT in curved space-time or in quantum
gravity, see e.g. \cite{SSS,Babic,Reuter,Bauer}. See also
\cite{CCvariable,Overduin} for various phenomenological models
with variable cosmological parameters. }. On the other hand a most
common possibility for $X$ in QFT and string theory would be some
scalar field $\chi$ (e.g. moduli or dilaton
fields\,\cite{Barreiro}) that results from low-energy string
theory. In fact, the original cosmon field was a
pseudo-dilaton\,\cite{PSW}. Here we will also use the ``cosmon''
denomination for $X$. This entity will stand for dynamical
contributions that go into the DE other than the vacuum energy
effects encoded in $\CC$. Since some of the effects associated to
$X$ could be linked to dynamical scalar fields, this justifies
its denomination of cosmon\,\cite{PSW}. Whatever it be its
ultimate nature, we require that the total DE density $\rD$ must
be conserved. We will show that the resulting $\CC$XCDM model can
provide a good qualitative description of the present data,
particularly the fitted EOS, and at the same time it helps to
tackle the cosmological coincidence problem.

The structure of the paper is as follows. In the next section we
describe a few general features of composite DE models and
construct the $\CC$XCDM model. In Section \ref{sect:solving1} we
explicitly solve the model. In Section \ref{sect:nucleosynthesis}
we consider the constraints imposed by nucleosynthesis and the
evolution of the dark energy. In Section \ref{sect:solving2} we
describe possible cosmological scenarios within the $\CC$XCDM
model. In Sect. \ref{sect:crossing} we explore the possibility of
crossing the $\we=-1$ divide. In Sect. \ref{sect:numerics} we
perform a comprehensive numerical analysis of the $\CC$XCDM model
and discuss how this model can help to alleviate the cosmological
coincidence problem. In the last section we present the final
discussion and we draw our conclusions.

\section{$\CC$XCDM model: dynamical DE made out of $\CC$ and cosmon.}
\label{sect:Lambdacosmon}

The simplest notion of dynamical dark energy can be implemented in
terms of a single dynamical field $X$ that completely supersedes
the cosmological constant. This point of view gave rise to the
notion of XCDM model\,\cite{TurnerWhite} as a substitute for the
standard $\CC$CDM one\,\cite{Peebles84}. When considering a XCDM
model with a single DE fluid of general kind (even a phantom
one), and under the general assumption $0<\OMo=\rMo/\rco<1$, the
energy density $\rD$ must be positive at present, $\rDo>0$,
otherwise it would be impossible to fulfill the cosmic sum rule
for flat universes: $\OMo+\ODo=1$, with $\ODo=\rDo/\rco$.
Furthermore, one usually assumes that this situation is always
the case, i.e. $\rD>0$ at any time in the cosmic evolution. This
entails that only the DE pressure, $\pD$, can be negative. For
example, for a quintessence field we have $\rD>0$ and
$-\rD<\pD<-\rD/3$, so that the strong energy condition (SEC) is
violated, but not the weak (WEC) and dominant (DEC) ones (cf.
Fig.\,\ref{fig1}). However, it should be noted that the data is
usually described by an overall effective EOS, $\pD=\we\rD$,
irrespective of the potential composite nature of the fluids that
build up the total DE density. In composite DE models, with a
mixture of fluids of individual EOS's\, $p_i=\omega_i\,\rho_i\
(i=1,2,...,n)$, the effective EOS of the mixture reads
\begin{equation}\label{mixture}
\we=\frac{\pD}{\rD}=\frac{\omega_1\,\rho_1+\omega_2\,\rho_2+...}{\rho_1+\rho_2+...}\,.
\end{equation}
Clearly, $\we$ is not related in a simple way to the various
$\omega_i$ and in general $\we=\we(z)$ will be a complicated
function of time or redshift even if all $\omega_i$ are constant.
A mixture of barotropic fluids is, therefore, non-barotropic in
general. As a consequence many possibilities open up. For example,
it should be perfectly possible in principle to have a mixture of
fluids in which one or more of them have negative energy density,
$\rho_i<0$, and positive pressure $p_i>0$, even at the present
time. These DE components can be phantom-like, in the sense that
may result in $\omega_i<-1$, but they are not phantoms of the
``standard'' type, for they violate the WEC and DEC while they can
still preserve the SEC. Such ``singular'' phantom components are
very peculiar because they actually reinforce the cosmological
gravitational pull, in addition to the one exerted by ordinary
matter (see Eq.\,(\ref{dda}) below). In this sense, rather than
acting as conventional phantom energy, these components behave as
a sort of unclustered ``matter'' (with negative energy) that may
be called ``Phantom Matter'' (PM) (cf. Fig.\,\ref{fig1}). Although
the latter is a bit bizarre, the rationale for it will appear
later. The point is that these components may eventually cause the
halt of the expansion. Phantom Matter could be an unexpected
component of DE, and one that still preserves the SEC as normal
matter does. In contrast, a standard phantom energy component
realizes the condition $\omega_i<-1$ in the form $p_i<-\rho_i<0$
(see Fig.\,\ref{fig1}). It violates all classical energy
conditions (WEC, DEC, SEC, including the null and null-dominant
energy conditions\,\cite{HawkingEOS}) and produces a specially
acute anti-gravitational effect (much stronger than that of a
positive cosmological term) which leads eventually to the
so-called ``Big Rip'' singularity, namely the Universe is ripped
to pieces after a super-accelerated phase of finite
duration\,\cite{Phantom}. Therefore, the two forms of
phantom-like behavior $\we<-1$ are dramatically different:
whereas standard phantom energy produces super-acceleration of
the expansion and leads to Big Rip destruction of the Universe,
Phantom Matter produces super-deceleration and fast stopping
(with subsequent reversal) of the expansion. That said, it is
obvious that what really matters in composite DE models is the
behavior of the overall parameters $(\rD,\pD)$. Irrespective of
the particular nature of the DE components we can still have e.g.
the overall density $\rD$ positive and the total effective $\pD$
negative. Then if $-1<\we<-1/3$ the effective behavior of the DE
fluid mixture is quintessence-like, while if $\we<-1$ it is
phantom-like (of the standard type), such that the cosmic sum
rule for flat universes can be preserved.

\FIGURE[t]{
\mbox{\resizebox*{0.45\textwidth}{!}{\includegraphics{figure1a.eps}}\
\ \ \ \ \
   \resizebox*{0.45\textwidth}{!}{\includegraphics{figure1b.eps}}}
\caption{Energy conditions:\ \textbf{(a)} The shaded regions
fulfill the Weak Energy Condition (WEC): $\rho\ge0$ and
$\rho+p\ge0$. The lighter-shaded one also satisfies the Dominant
Energy Condition (DEC): $\rho\ge|p|$; \textbf{(b)} The region
shaded in gray (with or without dots) fulfills the Strong Energy
Condition (SEC): $\rho+p\ge0$ and $\rho+3p\ge0$. The quintessence
(Q) region ($-1<\we<-1/3$) is marked cross-hatched, and the
\emph{usual} phantom region ($\we\le-1$ with $\rho>0$) is
indicated by P. The gray-dotted region corresponds to an unusual
kind of phantom $\we<-1$ with $\rho<0$, which we call ``Phantom
Matter'' (PM). Note that PM satisfies the SEC. The cosmon behaves
in many cases as PM within the $\CC$XCDM model.} \label{fig1} }

As we have mentioned, experimentally an effective (standard)
phantom fluid ($\rD>0$, $\we<-1$) for the DE cannot be discarded
at present because it is allowed by the combined analysis of the
supernovae and CMB data\,\cite{WMAP3Y,Jassal1,Jassal2}. In the
following we will construct a simple composite model where this
kind of mixture situations can be met. The resulting $\CC$XCDM
model is a kind of ``composite XCDM model'' because, in
contradistinction to the original XCDM\,\cite{TurnerWhite}, we
keep both the cosmological term (with a given status of dynamical
entity) and also the new dynamical fluid $X$, and we assume that
in general they are in interaction\,\footnote{One might entertain
whether one should better call it X$\CC$CDM, rather than
$\CC$XCDM. In the former case we would seem to imply that we
introduce $X$ to the $\CC$CDM model\,\cite{Peebles84}, whereas in
the latter we would be ``restoring'' $\CC$ into the $X$CDM
one\,\cite{TurnerWhite}. However, we emphasize that the presented
model is not just the result of addition, but also of interaction,
and in this sense the name should not be a problem. We stick to
$\CC$XCDM, with no subtle connotations whatsoever.}. Obviously a
negative value of the cosmological term at present ($\CC_0<0$) is
allowed in this framework (with the current \textit{total} DE
density $\rD>0$) thanks to the $X$ component. We will see that in
the future this may cause a halt of the expansion. This issue has
been investigated also in models with a fixed negative value of
the cosmological term\,\cite{Xinesos}. However, thanks to the
variability of $\CC$ and/or the PM behavior of $X$, the halt will
also be possible for $\CC_0>0$. The combined dynamics of $\CC$ and
cosmon $X$ opens many other possibilities that cannot be offered
by models which just add up a quintessence field to the standard
$\CC$CDM and commit to some particular form of the potential for
that field. The fact that we do not specify the nature of $X$ by
e.g. identifying it with a scalar field, endows our discussion
with a higher degree of generality.

{It should be stressed right from the start that despite the time
variation of the cosmological parameters within our framework,
this fact in no way contradicts the general covariance of the
theory. This is independent of whether such variation is
inherited from RG considerations or it has some general
unspecified origin\,\cite{CCvariable,Overduin}. The parameters
could even be variable in space as well (see e.g. \cite{SSS}),
although we will not allow them to do so here because we want to
preserve the Cosmological Principle. The general covariance is
insured by imposing the fulfillment of the general Bianchi
identity on both sides of Einstein's equations. The kind of
particular relations that this identity will imply among the
various cosmological quantities will depend on the metric itself,
which in our case we choose to be the
Friedmann-Lema\^{i}tre-Robertson-Walker (FLRW) one\,\footnote{A
different and certainly independent issue is whether these models
admit in general a Lagrangian formulation, and in this sense they
are phenomenological insofar as the connection to that
formulation is not straightforward. But in our opinion this
should not necessarily be considered a priori as a shortcoming.
After all the original introduction of the field equations (by
Einstein himself) was based purely on general covariance
considerations and with no reference whatsoever to an action
principle. This said, a Lagrangian formulation of some of these
models is not excluded and may lead to different phenomenological
consequences. From our point of view this issue must be decided
by experiment. For instance, in Ref.\,\cite{RGTypeIa} it was
shown in great detail how to test one of these RG models using
the distant supernovae data. }}. The most general form of the
Bianchi identity expressing the equation of continuity for a
mixture of fluids (\ref{mixture}) and variable cosmological
parameters reads:
\begin{equation}\label{generalBD}
\frac{d}{dt}\left[G\,\left(\sum_i\rho_i\right)\right]+G\,H\,\sum_i
\alpha_i\,\rho_i=0\,,\ \ \ \ \ \alpha_i\equiv 3(1+\omega_i)\,.
\end{equation}
In our case we assume that we have matter-radiation,
$\rmr=\rM+\rR$, and two DE components: one is called $X$
(``cosmon''), with dynamical density $\rX=\rX(t)$, and the other
is a cosmological term with density $\rL$, which can be constant
but in general it is also allowed to vary with the evolution:
$\rL=\rL(t)$. This is perfectly possible\,\cite{Peebles84} and
preserves the Cosmological Principle provided it only varies with
the cosmic time or (equivalently) with the cosmological redshift:
$\rL=\rL(z)$. The two DE components have barotropic indices $\wX$
and $\omega_{\CC}=-1$ respectively. In general the index $\wX$
for the cosmon will be dynamical, but in practice one can assume
that $X$ is some barotropic fluid with constant $\wX$, typically
in one of the two expected ranges $\wX\gtrsim -1$
(quintessence-like) and $\wX\lesssim -1$ (phantom-like). For the
moment we assume it is an arbitrary function of time or
redshift.  It should be clear that for the ``$\CC$ fluid'' the
relation $\omega_{\CC}=-1$ holds irrespective of whether $\rL$ is
strictly constant or variable. In this work we will keep the
gravitational coupling $G$ constant and assume that the
matter-radiation density $\rmr=\rM+\rR$ is conserved. Hence
Eq.\,(\ref{generalBD}) splits into corresponding conservation
laws for matter-radiation and total DE, namely
\begin{equation}\label{conslawrho}
\dot{\rho}_m+\,\amr\,\rmr\,H=0\,, \ \ \ \ \ \ \amr\equiv
3(1+\wm)\,,
\end{equation}
with
\begin{equation}\label{wm}
\wm=\frac13\,\frac{\rR}{\rM+\rR}\,,
\end{equation}
and
\begin{equation}\label{conslawDE}
\dot{\rho}_D+\aef\,\rD\,H=0\,, \ \ \ \ \ \ \aef\equiv
3\,(1+\we)\,,
\end{equation}
with
\begin{equation}\label{rD}
\rD=\rL+\rX.
\end{equation}
In Eq.(\ref{wm}) $\wm=0,1/3\ (\amr=3,4)$ for the matter dominated
epoch and the radiation dominated epoch respectively. Clearly,
the existence of a dynamical DE component $X$ is a necessary
condition for $\rL$ being variable in this framework. In the
particular case where $\rL$ is constant, $X$ would be
self-conserved, but in general $\rL$ and $X$ may interact and
therefore they can constitute two different dynamical components
of the total DE density (\ref{rD}). The effective barotropic index
-- or effective EOS parameter (\ref{mixture}) -- of the $\CC$XCDM
model is
\begin{equation}\label{eEOS}
\we=\frac{\pL+\pX}{\rL+\rX}=\frac{-\rL+\wX\,\rX}{\rL+\rX}=
-1+(1+\wX)\,\frac{\rX}{\rD}\,.
\end{equation}
As noticed above, even if $X$ is a barotropic fluid with constant
$\wX$ the mixture of $X$ and $\CC$ is non-barotropic because
$\rX$ and  $\rL$ will in general be functions of time or redshift,
and so will be $\we=\we(z)$. We shall compute this function
precisely for the $\CC$XCDM model under certain dynamical
assumptions on the behavior of the cosmological term. From the
equations above it is easy to see that the overall DE conservation
law (\ref{conslawDE}) can be rewritten as follows:
\begin{equation}\label{conslawDE2}
\dot{\rho}_{\Lambda}+\dot{\rho}_X+\,\aX\,\rX\,H=0\,, \ \ \ \ \
\aX\equiv 3(1+\wX)\,.
\end{equation}
For constant $\rL$, this law boils down to the self-conservation
of the $X$ component,
\begin{equation}\label{conslawX}
\dot{\rho}_X+\,\aX\,\rX\,H=0\,.
\end{equation}
For variable $\rL$, Eq.\,(\ref{conslawDE2}) shows that the
dynamics of the cosmological term and the cosmon become
entangled. One can confirm Eq.\,(\ref{conslawDE2}) starting from
the total energy-momentum tensor of the mixed DE fluid, with
$4$-velocity $U_{\mu}$:
\begin{equation}\label{TDE}
T^D_{\mu\nu}=T^{\CC}_{\mu\nu}+T^X_{\mu\nu}=
(\rL-\wX\,\rX)\,g_{\mu\nu}+(1+\wX)\rX\,U_{\mu}\,U_{\nu}\,.
\end{equation}
Then one can use the FLRW metric
\begin{equation}\label{FLRWm}
  ds^2=dt^2-a^2(t)\left(\frac{dr^2}{1-k\,r^2}
+r^2\,d\theta^2+r^2\,\sin^2\theta\,d\phi^2\right)
\end{equation}
and straightforwardly compute
$\bigtriangledown^{\mu}\,{T^D}_{\mu\nu}=0$. The result of course
is (\ref{conslawDE2}). Another fundamental equation of the
cosmological model under construction is Friedmann's equation
\begin{equation}
H^{2}\equiv \left( \frac{\dot{a}}{a}\right) ^{2}=\frac{8\pi\,G }{3}%
\left( \rmr +\rD\right) -\frac{k}{a^{2}}=\frac{8\pi\,G }{3}%
\left( \rmr +\rL+\rX\right) -\frac{k}{a^{2}}\,.  \label{FL}
\end{equation}
Let us also quote the dynamical field equation for the scale
factor:
\begin{equation}
\frac{\ddot{a}}{a}=-\frac
{4\pi\,G}{3}\,[\rmr\,(1+3\wm)+\rD\,(1+3\we)]=-\frac
{4\pi\,G}{3}\,[\rmr\,(1+3\wm)-2\,\rL+\rX\,(1+3\wX)]\,. \label{dda}
\end{equation}
This one is not independent from (\ref{FL}) and
(\ref{conslawDE2}), as can be easily seen by computing the time
derivative of $H$ with the help of equations
(\ref{FL}),(\ref{conslawDE}) and (\ref{conslawrho}):
\begin{equation}\label{dH}
\dot{H}=-\frac{4\pi\,G}{3}[\amr\,\rmr+\aef\rD]+\frac{k}{a^2}=
-\frac{4\pi\,G}{3}\,[\amr\,\rmr+\aX\rX]+\frac{k}{a^2}\,.
\end{equation}
Then Eq.(\ref{dda}) immediately follows from
${\ddot{a}}/{a}=\dot{H}+H^2$, showing that indeed it is not
independent. However it is useful to have (\ref{dda}) at hand
because it exhibits the acceleration law for the scale factor
under the combined influence of the matter-radiation density
$\rmr$ and the composite DE density (\ref{rD}).

How do we proceed next? Clearly, even if the index $\wX$ for the
cosmon would be given, we have three independent equations, e.g.
(\ref{conslawrho}),(\ref{conslawDE2}) and (\ref{FL}), but four
unknown functions $(H(t),\rmr(t),\rL(t),\rX(t))$. As a matter of
fact, Eq.(\ref{conslawrho}) is decoupled from the rest, and its
solution in the matter dominated and radiation dominated eras can
be cast in the simple unified form
\begin{equation}\label{rho}
\rmr(z)=\rmr^0\left(1+z\right)^{\amr}\,,
\end{equation}
where $\amr=3,4$ respectively for each era. For convenience we
have expressed this solution in terms of the cosmological
redshift, $z$, with the help of $H\,dt=-dz/(1+z)$. Notice that
Eq.(\ref{conslawDE}) can also be solved explicitly for the total
DE density:
\begin{eqnarray}\label{rDz}
\rD(z)=\rDo\,\exp\left\{3\,\int_0^z\,dz'
\frac{1+\we(z')}{1+z'}\right\}\,.
\end{eqnarray}
This equation is basically formal. In practice it cannot be used
unless the model is explicitly solved, meaning that we need first
to find the individual functions $\rL=\rL(z)$ and $\rX=\rX(z)$ for
the two components of the DE. In particular, this equation does
\textit{not} imply that $\rD=\rD(z)$ is positive definite for all
$z$ because, after solving the model, the function $\we=\we(z)$
may present singularities at certain values of $z$. The reciprocal
relation expressing the effective EOS parameter in terms of the
total DE density is
\begin{equation}\label{werD}
\we(z)=-1+\frac{1+z}{3}\,\frac{1}{\rD}\,\frac{d\rD}{dz}\,.
\end{equation}
From (\ref{eEOS}) and (\ref{werD}) we get
\begin{equation}\label{wXrX}
\aX\,\rX=\,(1+z)\,\frac{d\rD}{dz}
\end{equation}
which shows that the effective quintessence ($d\rD/dz>0$) or
phantom-like ($d\rD/dz<0$) character of the $\CC$XCDM model will
be known once the sign of $\aX$ \textit{and} the sign of the
density $\rX$ of the cosmon field are known. In particular, we
note the curious fact that even if $\wX<-1$ the model could still
effectively look as quintessence-like provided we admit the
possibility that the cosmon field can have negative energy-density
($\rX<0$)\,\footnote{The possibility of having negative energy
components has been entertained previously in the literature
within e.g. the context of constructing asymptotically de Sitter
phantom cosmologies\,\cite{McInness}.}. In general the $\CC$XCDM
model effectively behaves as quintessence (resp. phantom DE) if
and only if $\aX$ \textit{and} $\rX$ have the same (resp.
opposite) signs.

In the end we see that we have two coupled equations
(\ref{conslawDE2}) and (\ref{FL}) for three unknowns
$(H(t),\rL(t),\rX(t))$. We need a third equation to solve the
$\CC$XCDM model. Here we could provide either a specific model for
$X$ or one for $\rL$. Since $X$ represents a generic form of
dynamics other than the cosmological term (the scalar fields being
just one among many possible options), from our point of view it
is more fundamental to provide a model for $\rL$. The simplest
possibility is to assume $\rL=const.$ In this case the new model
is a trivial extension of the standard $\CC$CDM one, but even in
this case it leads to interesting new physics (see scenario I,
Section \ref{sect:solving2}). However, in the general case $\rL$
could be a dynamical quantity and a most appealing possibility is
to link its variability to the renormalization group
(RG)\,\cite{JHEPCC1}. Once the general principles of QFT in curved
space-time (or in Quantum Gravity\,\cite{Reuter}) will determine
the RG scaling law for $\CC$, the dynamics of $X$ will be tied to
the conservation law (\ref{conslawDE2}) in a completely
model-independent way; that is to say, independent of the ultimate
nature of the cosmon entity. This will be true provided we adhere
to the commonly accepted Ansatz that the total DE must be a
conserved quantity independent of matter. While we do not know for
the moment what are the general principles of QFT in connection to
the running properties of the cosmological term in Einstein's
equations, we have already some hints which may be put into test.
As a guide we will use the cosmological RG model described
in\,\cite{RGTypeIa}, based on the framework of
\,\cite{JHEPCC1,Babic} within QFT in curved space-time. A more
general RG cosmological model with both running $G$ and running
$\rL$ can also be constructed within QFT in curved space-time, see
Ref.\cite{SSS}. However, for simplicity hereafter we limit
ourselves to the case $G=const.$ Briefly, the model we will use is
based on a RG equation for the CT of the general form
\begin{equation}\label{RGEG1a}
(4\,\pi)^2\,\frac{d\rL}{d\ln\mu}=
\sum_{n=1}^{\infty}\,A_n\,\mu^{2n}\,.
\end{equation}
Here $\mu$ is the energy scale associated to the RG running in
cosmology. It has been argued that $\mu$ can be identified with
the Hubble parameter $\mu=H$ at any given epoch\,\cite{JHEPCC1}.
Since $H$ evolves with the cosmic time, the cosmological term
inherits a time-dependence through its primary scale evolution
with the renormalization scale $\mu$. Coefficients $A_{n}$ are
obtained after summing over the loop contributions of fields of
different masses $M_i$ and spins $\sigma_i$. The general behavior
is $A_n\sim \sum M_i^{4-2n}$\,\,\cite{JHEPCC1,Babic}. Therefore,
for $\mu\ll M_i$, the series above is an expansion in powers of
the small quantities $\mu/M_i$. Given that $A_{1}\sim\sum M_i^2$,
the heaviest fields give the dominant contribution. This trait
(``soft-decoupling'') represents a generalization of the
decoupling theorem in QFT-- see \cite{RGTypeIa} for a more
detailed discussion. Now, since $\mu=H_0\sim 10^{-33}\,eV$ the
condition $\mu\ll M_i$ is amply met for all known particles, and
the series on the \textit{r.h.s} of Eq.\,(\ref{RGEG1a}) converges
extremely fast. Only even powers of $\mu=H$ are consistent with
general covariance\,\cite{RGTypeIa}, barring potential bulk
viscosity effects in the cosmic fluid\,\cite{bulkvisco}. The
$n=0$ contribution is absent because it corresponds to terms
$\propto M_i^4$ that give an extremely fast evolution. Actually
from the RG point of view they are excluded because, as noted
above, $\mu\ll M_i$ for all known masses. In practice only the
first term $n=1$ is needed, with $M_i$ of the order of the
highest mass available. We may assume that the dominant masses
$M_i$ are of order of a high mass scale near the Planck mass
$M_P$. Let us define an effective mass $M$ as the total mass of
the heavy particles contributing to the $\beta$-function on the
\textit{r.h.s.} of (\ref{RGEG1a}) after taking into account their
multiplicities $a_i$\,\cite{RGTypeIa}:
\begin{equation}\label{Mdef}
M\equiv\sqrt{\left|\sum_i\,a_i\,M_i^2\right|}\,.
\end{equation}
We introduce (as in \cite{RGTypeIa}) the ratio
\begin{equation}\label{nu}
\nu=\frac{\sigma}{12\pi}\frac{M^2}{M_P^2}\,.
\end{equation}
Here $\sigma=\pm 1$ depending on whether bosons or fermions
dominate in their loop contributions to (\ref{RGEG1a}). Thus, if
the effective mass $M$ of the heavy particles is just $M_P$, then
$\nu$ takes the canonical value $\sigma\,\nu_0$, with
\begin{equation}\label{nu0}
\nu_0\equiv\frac{1}{12\,\pi}\simeq  2.6\times 10^{-2}\,.
\end{equation}
We cannot exclude a priori that $\nu$ can take values above
(\ref{nu0}) if these multiplicities are very high. Under the very
good approximation $n=1$, and with $\mu=H$, Eq.\,(\ref{RGEG1a})
abridges to\,\cite{RGTypeIa}
\begin{equation}\label{RGEG1b}
\frac{d\rL}{d\ln H}=\frac{3\,\nu}{4\,\pi}\,M_P^2\,H^2\,.
\end{equation}
 Equation (\ref{RGEG1b}) is the sought-for third equation needed to solve
our cosmological model. Its solution is
\begin{equation}\label{CCH}
\rL=c_0+c_1\,H^2\,,\\
\end{equation}
with
\begin{equation}\label{C0C1}
c_0=\rLo-\frac{3\,\nu}{8\pi}M_P^2\,H_0^2\,, \ \ \
c_1=\frac{3\,\nu}{8\pi}\,M_P^2\,,
\end{equation}
where $H=H(t)$ is given by (\ref{FL}). Notice that this result
gives some theoretical basis to the original investigation of the
possibility that $\CC\sim H^2$ (equivalently, $\rL\sim
H^2\,M_P^2$)\,\cite{Peebles84} at the present epoch. However, in
our case rather than a law of the type $\rL\sim H^2\,M_P^2$ we
have $\delta\rL\sim H^2\,M_P^2$, and the variation $\delta\rL$
around the reference value is controlled by a RG equation.

\section{Analytical solution of the $\CC$XCDM  model} \label{sect:solving1}

Using (\ref{conslawDE2}) (\ref{FL}), (\ref{dH}) and (\ref{RGEG1b})
we can present the relevant set of equations of the $\CC$XCDM
model in terms of the cosmological redshift variable $z$ as
follows:
\begin{eqnarray}\label{seteq}
&&\frac{d\,\rX}{dz}+\frac{d\,\rL}{dz}=\frac{\aX\,\rX}{1+z}\,,\nonumber\\
&&\frac{d\rL}{dz}=\frac{3\,\nu}{8\,\pi}\,M_P^2\,\frac{dH^2}{dz}\,,\nonumber\\
&&
\,\frac{dH^2}{dz}=\frac{8\,\pi\,G}{3}\,\frac{\amr\,\rmr+\aX\,\rX}{1+z}+2\,H_0^2\,\OKo\,(1+z)\,.
\end{eqnarray}
The latter can be replaced by Friedmann's equation (\ref{FL}) at
convenience. Here $\OKo=-k/H_0^2$ is the space curvature term at
the present time.

Equations (\ref{seteq}) must be solved for $(H,\rX,\rL)$, where
$\rmr$ is known from Eq.\,(\ref{rho}). Instead of $\rL,\rX$ we
will use
\begin{equation}\label{Omegas1}
\OL(z)=\frac{\rL(z)}{\rco}\,, \ \ \ \ \ \ \OX=\frac{\rX(z)}{\rco},
\end{equation}
where $\rco=3H^2_0/8\pi\,G$ is the critical density at present. We
also define
\begin{equation}\label{Omegas2}
\Om(z)=\frac{\rmr(z)}{\rco}\,,\ \ \ \ \OK(z)=\OKo\,(1+z)^2\,.
\end{equation}
From these equations we can derive the following differential
equation for $\OX$:
\begin{equation}\label{ODErX}
\frac{d\OX}{dz}-(1-\nu)\,\frac{\aX\,\OX}{1+z}=
-\nu\,\frac{\amr\,\Om}{1+z}-2\,\nu\,\OKo\,(1+z)\,.
\end{equation}
We can solve it for $\OX$ exactly, even if $\wX$ (hence $\aX$) is
an arbitrary function of the redshift. The result is the
following:
\begin{equation}\label{rXsolution}
\OX(z)=\OXo\,\left[1-\int_0^{z}\,dz''\,q(z'')\,\exp{\left\{-\int_0^{z''}\,dz'\,p(z')\right\}}\right]\,
\exp{\left\{\int_0^{z}\,dz'\,p(z')\right\}}\,,
\end{equation}
where
\begin{eqnarray}\label{pq}
&&p(z)=\frac{\aX\,(1-\nu)}{1+z}\,,\nonumber\\
&&q(z)=\nu\,\left[\amr\,\frac{\Omo}{\OXo}\,(1+z)^{\amr-1}+2\,\frac{\OKo}{\OXo}\,(1+z)\right]\,.
\end{eqnarray}
Notice that if $\nu=0$ then $\OL$ is strictly constant,
Eq.\,(\ref{ODErX}) reduces to Eq.\,(\ref{conslawX}) and $\OX(z)$
becomes self-conserved, as expected. If, in addition,
$\wX=$const. then (\ref{rXsolution}) collapses at once to
\begin{equation}\label{Oxnu0}
\OX(z)=\OXo\,(1+z)^{\aX}\,,
\end{equation}
which is of course the solution of (\ref{conslawX}) for constant
$\aX$. This situation is the simplest possibility, and we shall
further discuss it in Section \ref{sect:solving2} as a particular
case of our more general framework in which the two components of
the DE are dynamical and interacting with one another. The next to
simplest situation appears when $\nu\neq 0$ (still with
$\wX=$const.) because then there is an interaction between the two
components of the DE and these components can be treated as
barotropic fluids. In such case we can work out the integrals in
(\ref{rXsolution}) and obtain a non-trivial fully analytical
solution of the system (\ref{seteq}) for arbitrary spatial
curvature. The final result can be cast as follows:
\begin{eqnarray}\label{solved1}
&&  \frac{H^2(z)}{H_0^2}=\frac{\OLo-\nu}{1-\nu}+F(z)\,,\nonumber\\
&&  \OL(z)=\frac{\OLo-\nu}{1-\nu}+\nu\,F(z)\,,\nonumber\\
&&\OX(z)=(1-\nu)\,F(z)-\Omo\,(1+z)^{\amr}-\OKo\,(1+z)^2\,,
\end{eqnarray}
with
\begin{eqnarray}\label{Fz}
F(z)=&&\frac{\Omo(\amr-\aX)}{\amr-\aX\,(1-\nu)}\,(1+z)^{\amr}
-\frac{\,\OKo(\aX-2)}{2-\aX\,(1-\nu)}\,(1+z)^2\nonumber\\
&&
+\left[\frac{1-\OLo}{1-\nu}-\frac{\Omo(\amr-\aX)}{\amr-\aX\,(1-\nu)}+
\frac{\,\OKo(\aX-2)}{2-\aX\,(1-\nu)}\right]\,(1+z)^{\aX\,(1-\nu)}\,.
\end{eqnarray}
The solution above corresponds to the matter ($\amr=3$) or
radiation ($\amr=4$) dominated eras. The expansion rate for the
$\CC$XCDM model in any of these eras can be expressed as
\begin{equation}\label{FL2}
H^2(z)=H_0^2\,\left[\Omo\,(1+z)^{\amr}+\OKo\,(1+z)^2+\OD(z)\,\right]\,,
\end{equation}
where $\OD(z)=\rD(z)/\rco$ is the total DE energy density in units
of the current critical density. From the previous formulae the
latter can be written in full as follows:
\begin{eqnarray}\label{ODz}
\OD(z)=&&\frac{\OLo-\nu}{1-\nu}-\frac{\nu\,\aX\,\Omo\,(1+z)^{\amr}}{\amr-\aX\,(1-\nu)}-
\frac{\nu\,\aX\,\OKo\,(1+z)^2}{2-\aX\,(1-\nu)}\nonumber\\
&&+\left[\frac{1-\OLo}{1-\nu}-\frac{\Omo(\amr-\aX)}{\amr-\aX\,(1-\nu)}+
\frac{\,\OKo(\aX-2)}{2-\aX\,(1-\nu)}\right]
\,(1+z)^{\aX\,(1-\nu)}\,.
\end{eqnarray}
Let us finally derive the generalized form of the ``cosmic sum
rule'' within the $\CC$XCDM model. From (\ref{ODz}) we can readily
verify that for $z=0$ we have $\OD(0)=1-\OMo-\OKo$. This renders
the desired sum rule satisfied by the current values of the
cosmological parameters in the $\CC$XCDM:
\begin{equation}\label{sumrule0}
\OMo+\ODo+\OKo=\OMo+\OLo+\OXo+\OKo=1\,.
\end{equation}
The cosmic sum rule can be extended for any redshift
\begin{equation}\label{sumrule1}
\tilde{\Omega}_m(z)+\tilde{\Omega}_{\Lambda}(z)+\tilde{\Omega}_X(z)+\tilde{\Omega}_K(z)=1\,,
\end{equation}
provided we define the new cosmological functions
\begin{equation}\label{Omegas3}
\tilde{\Omega}_m(z)=\frac{\rmr(z)}{\rc(z)}\,,\ \
\tilde{\Omega}_{\Lambda}(z)=\frac{\rL(z)}{\rc(z)}\,, \ \
\tilde{\Omega}_X(z)=\frac{\rX(z)}{\rc(z)},\ \
\tilde{\Omega}_K(z)=-\frac{k}{H^2(z)}\,(1+z)^2\,,
\end{equation}
with $\rc(z)=3H^2(z)/8\pi\,G$ the critical density at redshift
$z$. These functions should not be confused with (\ref{Omegas1})
and (\ref{Omegas2}). We will, however, use more frequently the
latter.

We may ask ourselves: in which limit do we recover the standard
$\CC$CDM model? We first notice that for $\nu=0$ the function
(\ref{Fz}) dwarfs to
\begin{equation}\label{FLCDM}
F(z)=\Omo\,(1+z)^{\amr}+\OKo\,(1+z)^2+\OXo\,(1+z)^{\aX}\,.
\end{equation}
where we have used (\ref{sumrule0}). Then substituting this in
(\ref{solved1}) one can immediately check that the standard
$\CC$CDM model formulae are obtained from those of the $\CC$XCDM
model in the limit $\nu\rightarrow 0$ \textit{and}
$\OXo\rightarrow 0$, as could be expected.

\section{Nucleosynthesis constraints and the evolution of the DE from the radiation
era into the matter dominated epoch} \label{sect:nucleosynthesis}

At temperatures near $1\,MeV$ the weak interactions (responsible
for neutrons and protons to be in equilibrium) freeze-out. The
expansion rate is sensitive to the amount of DE, and therefore
primordial nucleosynthesis ($T\sim 0.1-1\,MeV$) can place
stringent bounds on the parameters of the $\CC$XCDM model. A
similar situation was considered in the running cosmological
models previously studied in Ref.\,\cite{RGTypeIa,SS1}. In the
latter, nucleosynthesis furnished a bound on the parameter $\nu$
defined in Eq.(\ref{nu}) from the fact that the ratio of the
cosmological constant density to the matter density should be
relatively small at the nucleosynthesis epoch, $(\rL/\rmr)_N\ll
1$, of order $10\%$ (see also \cite{Ferreira97}). This condition
led to $\nu<0.1$, which is e.g. amply satisfied by the natural
value (\ref{nu0}). However, in the $\CC$XCDM model the total DE
density is not just $\rL$, but $\rD$ given by (\ref{rD}), and
moreover the dynamics of the cosmological term here is not linked
to the matter density (as in Ref.\cite{RGTypeIa}) but to the
cosmon density. Therefore, the appropriate ratio to be defined in
the present instance is
\begin{equation}\label{ratio}
r\equiv\frac{\rD}{\rmr}=\frac{\rL+\rX}{\rmr}=\frac{\OL+\OX}{\Om}\,.
\end{equation}
Notice that for the flat case ($k=0$) this ratio is related to
$\tilde{\Omega}_D\equiv\tilde{\Omega}_{\Lambda}+\tilde{\Omega}_{X}$
(cf. Eq.(\ref{Omegas3})) as follows:
\begin{equation}\label{randrD}
\tilde{\Omega}_D=\frac{r}{1+r}\,.
\end{equation}
Even if $k\neq 0$, at the nucleosynthesis time the curvature term
is negligible. Therefore the nucleosynthesis bounds on
$\tilde{\Omega}_D$ of order $10\%$\,\cite{Ferreira97} are
essentially bounds on $r$. Let us compute the ratio $r$ in
general and then evaluate it at nucleosynthesis.  For convenience
let us use the function (\ref{Fz}) and express the result as
\begin{equation}\label{rz}
r(z)=\frac{\frac{\OLo-\nu}{1-\nu}+F(z)-\Omo\,(1+z)^{\amr}-\OKo\,(1+z)^{2}}{\Omo(1+z)^{\amr}}\,.
\end{equation}
It is easy to check that for $z=0$ this equation reduces to the
expected result
\begin{equation}\label{rz0}
r_0\equiv r(0)=\frac{\OLo+\OXo}{\OMo}=\frac{\ODo}{\OMo}\,.
\end{equation}
Let now $z_N\sim T_N/T_0\sim (0.1\,MeV)/(2\times 10^{-4}\,eV)\sim
10^9$ be the typical cosmological redshift at nucleosynthesis
time. Let also $\wR=1/3$ and $\aR=4$ be the values of the
parameters $\wm$ and $\amr$ at nucleosynthesis (and in general in
the radiation epoch). We can obviously neglect the first
(constant) term in the numerator of (\ref{rz}) at $z=z_N$, and as
we have said also the contribution from the spatial curvature
(which scales with two powers less of $(1+z)$ as compared to the
radiation density). After a straightforward calculation we obtain
\begin{equation}\label{rzN}
r_N\equiv
r(z_N)\simeq\frac{F(z_N)-\ORo\,(1+z_N)^{4}}{\ORo(1+z_N)^{4}}
 =\re+R_N(z_N)
\end{equation}
where
\begin{equation}\label{re}
\re\equiv -\frac{\epsilon}{\wR-\wX+\epsilon}\,,
\end{equation}
with
\begin{equation}\label{epsilon}
\epsilon\equiv \,\nu\,(1+\wX)\,,
\end{equation}
and we have defined the function
\begin{equation}\label{Rz}
R_N(z_N)=\left[\frac{1-\OLo}{\ORo(1-\nu)}-\frac{\wR-\wX}{\wR-\wX+\epsilon}-
\frac{\OKo}{\ORo}\,\frac{1+3\,\wX}{1+3\,\wX-3\,\epsilon}\right]\,(1+z_N)^{3\,(\wX-\epsilon)-1}\,.
\end{equation}
Here $\ORo$ is the radiation density at present. The parameter
$\epsilon$ defined in (\ref{epsilon}) will soon play an important
role in our discussion. The ``residual'' function (\ref{Rz}) will
be totally harmless at nucleosynthesis provided the exponent of
$(1+z_N)$ that goes with it is sufficiently large (in absolute
value) and negative, in which case $R_N(z_N\sim 10^9)$ will be
negligible. We start with the condition that it be negative, which
provides the constraint
\begin{equation}\label{wxe}
\wX-\epsilon<\frac13\,.
\end{equation}
However, there is also the first term on the \textit{r.h.s.} of
(\ref{rzN}). This term does not decay with the redshift, it is
constant and therefore must be small by itself. If we take the
same condition as in \cite{RGTypeIa} concerning the bound on the
ratio of DE density to radiation density at nucleosynthesis,
namely $|r_N|<10\%$, we find the additional constraint
\begin{equation}\label{nb1}
\left|\re\right|=\left|\frac{\epsilon}{\wR-\wX+\epsilon}\right|<0.1\,\,.
\end{equation}
We shall assume that the cosmon barotropic index lies in the
interval
\begin{equation}\label{wxrange}
-1-\delta/3<\wX<-1/3  \ \ \ \ (\delta>0)\,,
\end{equation}
with $\delta$ a small positive quantity. The upper limit is fixed
by requiring that $X$ has at most quintessence behavior (i.e.
excluding matter-radiation behavior); the lower limit is expected
to be some number below (but around) $-1$ in order to accommodate
a possible phantom-like behavior. From Eq.\,(\ref{wxrange}), and
recalling that $\wR=1/3$, it is patent that the bound (\ref{nb1})
just translates into a rough bound on $\epsilon$ itself:
\begin{equation}\label{eb}
|\re|\simeq|\epsilon|=|\nu\,(1+\wX)|<0.1\,.
\end{equation}
Using this relation and the condition (\ref{wxrange}) we see that
the previous bound (\ref{wxe}) is amply satisfied and the function
(\ref{Rz}) actually behaves as
\begin{equation}\label{aN}
R_N(z)\sim (1+z)^{\beta}\ \ \ \ \ (\beta<-2-3\,\epsilon)\,.
\end{equation}
This means that $R_N(z_N\sim 10^9)$ will be completely
negligible, as desired. Therefore, the two bounds from
nucleosynthesis just reduce to one, Eq.\,(\ref{nb1}), and we can
simply set there $|r_N|=|\re|$. We will use the exact bound
(\ref{nb1}) for our numerical analysis, but the approximate
version (\ref{eb}) is useful for qualitative discussions.

In contrast to the nucleosynthesis bound obtained in
Ref.\,\cite{RGTypeIa} for the alternative scenario where the DE
was composed only of a running cosmological constant (in
interaction with matter) we see that for the $\CC$XCDM model it
is not necessary to require that $|\nu|$ is small, but only the
product of $|\nu|$ times $|\aX|$.  We also note that in the
$\CC$XCDM model for $\nu\neq 0$ there is an irreducible
contribution of the DE to the total energy density in the
radiation era, whose size is basically fixed by the tolerance of
the nucleosynthesis processes to the presence of a certain amount
of dark energy. The existence of this residual constant
contribution shows that in the $\CC$XCDM model the presence of
the DE component takes place at all epochs of the Universe
evolution. It is therefore interesting to study what will be the
evolution of this tiny (but not necessarily negligible)
contribution in later times, in particular in the matter
dominated epoch, where $\wm=0$ ($\amr=3$). Let us thus compute
the ratio (\ref{ratio}) in this epoch, and call it $r_M(z)$. In
this case we apply the full formula (\ref{rz}), without
neglecting any term, because $z$ in the matter-dominated epoch
covers a range from $\sim 10^3$ to (near) $-1$. The final result
is:
\begin{equation}\label{rzMD}
r_M(z)
=\frac{\OLo-\nu}{(1-\nu)\,\OMo\,(1+z)^3}+\frac{\OKo}{\OMo}\,
\frac{3\,\epsilon}{(1+3\,\wX-3\,\epsilon)\,(1+z)}+\frac{\epsilon}{\wX-\epsilon}+R_M(z)\,,
\end{equation}
with
\begin{equation}\label{RzM}
R_M(z)=\left[\frac{1-\OLo}{\OMo(1-\nu)}-\frac{\wX}{\wX-\epsilon}-
\frac{\OKo}{\OMo}\,\frac{1+3\,\wX}{1+3\,\wX-3\,\epsilon}\right]\,(1+z)^{3\,(\wX-\epsilon)}\,.
\end{equation}
It is easy to check that for $z=0$ the expression (\ref{rzMD})
shrinks just to (\ref{rz0}). For general $z$ we see that the
first term on the \textit{r.h.s.} of (\ref{rzMD}) goes like
$(1+z)^{-3}$ and hence evolves fast with the expansion and is
unbounded for $z\rightarrow -1$. The second term (the curvature
dependent term) evolves much more slowly, namely as $(1+z)^{-1}$
and it can be neglected to study the future evolution (also
because $\OKo$ is very small). The rising of $r(z)$ with the
expansion cannot be stopped in the standard $\CC$CDM model
because for it the ratio just reads
\begin{equation}\label{rzLCDM}
r_M(z) =\frac{\OLo}{\,\OMo\,(1+z)^3}\,.
\end{equation}
This particular result is of course recovered from (\ref{rzMD}) in
the limit $\nu=0$ and $\OXo=0$ (see Section \ref{sect:solving1}).
However, in the $\CC$XCDM the presence of the non-trivial function
$R_M(z)\sim (1+z)^{3\,(\wX-\epsilon)}$ is crucial as it allows
the existence of both a stopping point, $z_s$, as well as of an
extremum of the function (\ref{rzMD}) at some $z=\zm$ during the
future Universe evolution. We will study the relationship between
these two features in the next section and we will see that the
stopping condition is correlated with this extremum being a
maximum. Here we will concentrate on the conditions of existence
of this maximum of $r_M(z)$.

Let us from now on restrict our considerations to the flat space
case, which is the most favored one. The redshift position $\zm$
of the maximum can be worked out from Eq.\,(\ref{rzMD}). The
result is the following:
\begin{equation}\label{zs}
1+\zm=\left[\frac{\OLo-\nu}{\wX\,(\OXo+\nu\,\OMo)-
\epsilon\,(1-\OLo)}\right]^{\frac{1}{3\,(1+\wX-\epsilon)}}\,.
\end{equation}
The nature of this extremum is obtained by working out the second
derivative of $r(z)$ at $z=\zm$. We find
\begin{equation}\label{rpp}
r''(\zm)=\frac{3\,\aX}{(1+\zm)^5}\,\frac{\OLo-\nu}{\OMo}\,,
\end{equation}
where $\zm$ is given by (\ref{zs}). A sufficient condition for the
extremum to be a maximum is that
\begin{equation}\label{maximum}
  \aX\,\left(\OLo-\nu\right)<0\,.
\end{equation}
The height  of the maximum, $r_{\rm max}\equiv r(\zm)$, is the
following:
\begin{equation}\label{height}
r_{\rm
max}=\frac{1}{(1+\zm)^3}\,\frac{\OLo-\nu}{\OMo}\,\frac{1+\wX}{\wX-\epsilon}+\frac{\epsilon}{\wX-\epsilon}\,,
\end{equation}
where the last term on the \text{r.h.s.} is negligible.

Notice that the aforementioned maximum can only exist in the
future (or at present), but not in the past. Its existence in the
past would be incompatible with the experimentally confirmed state
of acceleration of the Universe that started in relatively recent
times. To prove this, one can show that the acceleration
Eq.\,(\ref{dda}) can be written in terms of the ratio
(\ref{ratio}) and its derivative $r'(z)$ with respect to redshift
as follows. We first write Friedmann's equation for $k=0$ in the
form $H^2=({8\pi\,G}/{3})\,(1+r)\,\rmr$. Then we differentiate it
with respect to the cosmic time,
\begin{equation}\label{dH2}
\dot{H}=-\frac{4\pi\,G}{3}\left[(1+r)\,\amr-\frac{\dot{r}}{H}\right]\,\rmr\,,
\end{equation}
where use has been made of the equation of continuity
(\ref{conslawrho}). With the help of these equations we compute
${\ddot{a}}/{a}=\dot{H}+H^2$:
\begin{eqnarray}
\frac{\ddot{a}}{a}&=&-\frac
{4\pi\,G}{3}\,\left[(1+3\wm)(1+r(t))-\frac{\dot{r}(t)}{H})\right]\,\rmr\nonumber\\
&=&-\frac
{4\pi\,G}{3}\,\left[(1+3\wm)(1+r(z))+(1+z)\,r'(z)\right]\,\rmr\,,\label{dda2}
\end{eqnarray}
where $r'(z)=dr(z)/dz$. The last equation shows that if the
maximum could take place at some point $z=z_1>0$ in the recent
past, then any $z$ in the interval $0<z<z_1$ would satisfy
$r'(z)>0$, and then Eq.\,(\ref{dda2}) would immediately imply
$\ddot{a}<0$, which is incompatible with the known existence of
the acceleration period in our recent past (\textit{q.e.d.}).

Therefore, the maximum will in general lie ahead in time. After
surpassing $z=\zm$ into the future, the ratio (\ref{rzMD}) drops
fast (in redshift units) to $r_M(z_s)=-1$, where $z_s$ can be
relatively close to $-1$. At this point $\rD$ takes the negative
value $\rD(z_s)=-\rMo\,(1+z_s)^3$, with $|\rD(z_s)|\ll\rD^0\ $.
From the previous considerations it is clear that the maximum
(\ref{zs}) can only exist for
\begin{equation}\label{existmax}
\frac{\OLo-\nu}{\wX\,(\OXo+\nu\,\OMo)- \epsilon\,(1-\OLo)}>0\,.
\end{equation}
We point out that if the cosmon $X$ would behave as another pure
cosmological term (namely $\wX=-1$) then $\epsilon=0$ and the
maximum (\ref{zs}) would not exist. On the other hand, for $\nu=0$
then again $\epsilon=0$, and for $\wX\neq -1$ the maximum still
exists and is located at
\begin{equation}\label{zsnu0}
1+\zm=\left[\frac{\OLo}{\wX\,\OXo}\right]^{\frac{1}{\aX}}\ \ \ \
(\nu=0)\,.
\end{equation}
Since $\wX<0$ this enforces $\OLo$ and $\OXo$ to have opposite
signs.

The various scenarios will be further discussed in the next
section from a more general point of view, where we will link the
conditions for the stopping of the Universe's expansion with the
conditions of existence of the maximum of (\ref{rzMD}) before
reaching the turning point. The presence of these two points is a
central issue in our discussion of the cosmological coincidence
problem within the $\CC$XCDM . The numerical analysis displaying
these features is presented in Section \ref{sect:numerics}. But
before that, let us take a closer look to the analytical results
and their interpretation.

\section{Cosmological scenarios in the $\CC$XCDM model. Autonomous system.} \label{sect:solving2}

 For a better understanding of the solution
that we have found in Section \ref{sect:solving1} it will be
useful to look at the set of equations (\ref{seteq}) from the
point of view of the qualitative methods for systems of
differential equations. This should help in the interpretation of
the numerical analysis performed in Section \ref{sect:numerics}.
As a matter of fact, we will end up by solving quantitatively the
problem with this alternate method and use it as a cross-check.
Again it will simplify our analysis if we assume that the cosmon
is a barotropic fluid with $\wX=$ const. Let us introduce the
convenient variable
\begin{equation}\label{zetav}
\zeta=-\ln (1+z)\,.
\end{equation}
In this variable the remote past ($t=0$) lies at $\zeta=-\infty$
whereas the remote future ($t=\infty$) is at $\zeta=+\infty$. In
the previous section we have solved the $\CC$XCDM cosmological
model exactly for any value of the spatial curvature, but in
practice we will set the curvature parameter $\OKo$ to zero to
conform with the small value suggested by the present
observations\,\cite{WMAP03,WMAP3Y}. This will allow us to deal
with the basic set of equations (\ref{seteq}) as an autonomous
linear system of differential equations in the variable
(\ref{zetav}). Indeed, after a straightforward calculation we find
\begin{eqnarray}\label{autonomous1}
&&\frac{d\OX}{d\zeta}=-\left[\nu\,\amr+(1-\nu)\,\aX\right]\,\OX-\nu\,\amr\,\OL+\nu\,\amr\,\Oc\,,\nonumber\\
&&\frac{d\OL}{d\zeta}=\nu\,(\amr-\aX)\,\OX+\nu\,\amr\,\OL-\nu\,\amr\,\Oc\,,\nonumber\\
&& \,\frac{d\Oc}{d\zeta}=(\amr-\aX)\,\OX+\,\amr\,\OL-\amr\,\Oc\,,
\end{eqnarray}
where we have defined  $\Oc(z)=\rc(z)/\rco$. Notice that by
virtue of (\ref{sumrule1})
\begin{equation}\label{sumrule2}
\Om(z)+\OL(z)+\OX(z)=\Oc(z)\,.
\end{equation}
We have actually used this equation to eleminate $\Om$ from
(\ref{autonomous1}). The matrix of the system (\ref{autonomous1})
has zero determinant. The eigenvalues read
\begin{equation}\label{eigenvalues}
\lambda_1=-\aX\,(1-\nu)\,,\ \ \lambda_2=-\amr\,,\ \ \
\lambda_3=0\,.
\end{equation}
The corresponding eigenvectors (up to a constant) are
\begin{equation}\label{eigenvectors}
{\bf v_1}=\begin{pmatrix}
  1-\nu \\
  \nu \\
  1
\end{pmatrix}\,,\ \ \ {\bf v_2}=\begin{pmatrix}
  \frac{-\nu\,\amr}{\amr-\aX} \\
  \nu \\
  1
\end{pmatrix}\,,\ \ \ {\bf v_3}=\begin{pmatrix}
  0 \\
  1 \\
  1
\end{pmatrix}\,.
\end{equation}
If we define the column vector
\begin{equation}\label{Omegav}
{\bf \Omega}(\zeta)=\begin{pmatrix}
  \OX(\zeta) \\
  \OL(\zeta) \\
  \Oc(\zeta)
\end{pmatrix}
\end{equation}
then the solution of the autonomous system (\ref{autonomous1})
reads
\begin{equation}\label{solve2}
{\bf \Omega}(\zeta)=C_1\,{\bf v_1}\,e^{\lambda_1\,\zeta}+C_2\,{\bf
v_2}\,e^{\lambda_2\,\zeta}+C_3\,{\bf v_3}\,,
\end{equation}
with
\begin{equation}\label{Cs}
C_1=\frac{1-\OLo}{1-\nu}-\frac{\Omo(\amr-\aX)}{\amr-\aX\,(1-\nu)}\,,\
\ C_2=\frac{\Omo(\amr-\aX)}{\amr-\aX\,(1-\nu)}\,,\ \
C_3=\frac{\OLo-\nu}{1-\nu}\,.
\end{equation}
The values of these coefficients are fixed by the boundary
condition
\begin{equation}\label{boundary}
{\bf \Omega}(\zeta=0)=\begin{pmatrix}
  \OXo \\
  \OLo \\
  \Oco
\end{pmatrix}\,.
\end{equation}
One can check that the solution obtained by this method is
completely consistent with the one obtained in Section
\ref{sect:solving1} for the flat space case.

Once the solution is expressed in terms of the eigenvalues, as in
(\ref{solve2}),
 we can investigate the qualitative behaviour of
the dynamical system. Let us describe some possible scenarios for
the phase trajectories. Let us recall that the cosmon barotropic
index lies in the interval (\ref{wxrange}). Equivalently,
\begin{equation}\label{alfas}
-\delta<\aX<2<\amr\, \ \ \ \ \ \  (\delta>0)\,.
\end{equation}
The scenarios that we wish to consider are mainly the ones that
are compatible with the existence of a stopping point in the
future Universe expansion. The reason for this is that if there is
a stopping point then we can argue (see below) that it can provide
an explanation for the cosmic coincidence problem. Of course, this
``stopping point'' is actually a ``turning point'', meaning that
the Universe ceases its expansion there, but it subsequently
reverses its evolution (then contraction) towards the Big Crunch.
Among the possible cosmological scenarios let us remark the
following:
\begin{itemize}

\item I) $0<\aX<2$ and $\nu=0$. In this case the two non-vanishing eigenvalues are
both negative, $\lu<0\,,\ld<0$, meaning that there is an
asymptotically stable node. All phase trajectories of the system
in vector space (\ref{Omegav}) become attracted (as
$\zeta\rightarrow\infty$) towards the vector node
\begin{equation}\label{nu0node}
{\bf \Omega}^{*}_{I}=\begin{pmatrix}
  0 \\
  \OLo \\
  \OLo
\end{pmatrix}\,.
\end{equation}
For $\OLo>0$ this fixed point is obviously de Sitter space-time.
However, if we further impose the condition $\OLo<0$, the
convergence to that node will be stopped because the third
component of the vector solution (\ref{Omegav}),
$\Oc(z)=\rc(z)/\rco=H^2(z)/H_0^2$, cannot become negative. The
interpretation of this case is clear: we are in a situation where
we have a quintessence cosmon entity $X$ ($-1<\wX<-1/3$), with
density (\ref{Oxnu0}), superimposed on a strictly constant
cosmological term:
\begin{equation}\label{ODznu0}
\OD(z)=\OLo+ \OXo\,(1+z)^{\aX}\,.
\end{equation}
The requirement $\OLo<0$ eventually stops the Universe expansion
at some point in the future because the $X$ density gradually
diminishes with the cosmic time and the \textit{r.h.s.} of
(\ref{FL2}) becomes zero. We remark that the possibility that
$\OLo<0$ should not be discarded a priori. Even though it is
clearly unfavored by measurements within the standard $\CC$CDM
model\,\cite{Supernovae,WMAP03,LSS,WMAP3Y}, it is perfectly
possible within the $\CC$XCDM whose generalized cosmic sum rule
(\ref{sumrule0}) for flat space reads
\begin{equation}\label{sumrule0flat}
\OMo+\OLo+\OXo=1\,.
\end{equation}
For example, if we take the approximate prior $\OMo=0.3$
(typically from the LSS inventory of matter obtained from galaxy
clusters analysis\,\cite{LSS}), then a value $\OLo<0$ can be
easily compensated for by a positive cosmon density at present:
$\OXo>0$. Notice that the measurements of distant supernovae and
CMB should in principle be sensitive to the total DE content
$\OD=\OL+\OX$, and of course to the total matter content.
Moreover, being $\aX>0$ we see that $\OLo<0$ is compatible with
the existence of the maximum of $r(z)$ at $z=\zm$ -- cf.
Eq.(\ref{maximum}).

\item II) $-\delta<\aX<0$ and $\nu=0$. Here the non-vanishing eigenvalues are of
different sign: $\lu>0\,,\ld<0$. It follows that (\ref{nu0node})
becomes a saddle point and the phase trajectories depart from it
more and more with the evolution. Equation (\ref{ODznu0}) still
holds in this case, and for $\OXo>0$ the cosmon density increases
indefinitely with the expansion. In fact, $X$ behaves here as a
standard phantom field (cf. Fig.\ref{fig1}) and produces a
super-accelerated expansion beyond the normal de Sitter expansion.
In these conditions the Universe's evolution strips apart
gravitationally bound objects (and ultimately all bound systems),
i.e. it ends up in a frenzy ``Big Rip''\,\cite{Phantom}. However,
if $\OXo<0$ the \textit{r.h.s.} of (\ref{FL2}) eventually becomes
zero and the Universe stops its expansion at some point in the
future. We can also see these features in terms of
Eq.(\ref{solve2}). The sign of the coefficient $C_1$ determines
in this case whether there can be stopping ($C_1<0$) or not
($C_1>0$). Since $\nu=0$ here, we have $C_1=1-\OLo-\OMo=\OXo$ and
therefore it is the sign of $\OXo$ that controls the stopping
along the lines mentioned above. Notice that although $\wX<-1$,
the Big Rip is impossible for $\OXo<0$ because the term
$-\rX\,(1+3\wX)$ on the \textit{r.h.s.} of (\ref{dda}) is
negative, therefore the cosmon in this instance actually
collaborates with matter to enhance more and more the
cosmological deceleration until the Universe reaches a turning
point. In this case the cosmon $X$ acts as ``Phantom Matter''
(PM), the only component of the DE that still preserves the SEC
(see Fig\,1). We stress that even though $X$ behaves as PM here,
the sign of the product $\aX\,\rX$ is positive and hence the
effective behavior of the $\CC$XCDM model remains
quintessence-like -- see Eq.(\ref{wXrX}).

\item III) $\aX=0$ and $\nu\neq 0$.  Two eigenvalues are zero
($\lu=\lt=0$) and $\ld<0$. One can check that the following node
is the end point of all trajectories:
\begin{equation}\label{alfa0de}
{\bf \Omega}^{*}_{III}(\nu)=\begin{pmatrix}
  \OXo+\nu\,\OMo \\
  \OLo-\nu\,\OMo \\
  1-\OMo
\end{pmatrix}\,.
\end{equation}
This situation corresponds to effectively having two variable
``$\CC$'s'', one is $X$ (which has EOS parameter $\wX=-1$) and the
other is $\CC$. Both ``$\CC$'s'' are variable because $\nu\neq 0$,
but the sum $\rL(z)+\rX(z)=const.$ i.e. $\rD(z)$ behaves as a
strictly constant cosmological term -- cf.
Eq.\,(\ref{conslawDE2}). With $\OMo<1$, there is no possibility
for stopping the expansion under these circumstances, no matter
how $\CC$ varies, because $\Oc(z)$ (hence $H^2(z)$) remains always
positive. Although this case is uninteresting for our purposes, it
shows at least that there is no turning point in the Universe
evolution if the cosmon has the exact EOS of a cosmological term.
In Section \ref{sect:nucleosynthesis} we also proved that under
these circumstances there is no maximum of $r=\rD/\rmr\,$ either.
The Universe moves asymptotically towards an ever-expanding de
Sitter phase characterized by an effective cosmological constant
given by $\OLo+\OXo=1-\OMo>0$. We conclude that a necessary
condition for the existence of a turning point in the $\CC$XCDM
model is that the cosmon EOS parameter is different from $-1$; in
other words, $\aX=0$ is to be excluded from Eq.\,(\ref{alfas}).
This will be understood implicitly hereafter.

\item IV) $0<\aX<2$ and $\nu<1$ (including of course $\nu<0$).
As in case I above the two eigenvalues are both negative
$\lu<0\,,\ld<0$ and there is a ($\nu$-dependent) node onto which
all phase trajectories are attracted, namely
\begin{equation}\label{nunode}
{\bf \Omega}^{*}_{IV}(\nu)=\begin{pmatrix}
  0 \\
  \frac{\OLo-\nu}{1-\nu} \\
  \frac{\OLo-\nu}{1-\nu}
\end{pmatrix}\,.
\end{equation}
For $\nu=0$, ${\bf \Omega}^{*}_{IV}(0)={\bf \Omega}^{*}_{I}$ and
we recover the case I above. Again the attraction towards the
node can be stopped under suitable conditions. In this case we
have to require that
\begin{equation}\label{cond1}
\frac{\OLo-\nu}{1-\nu}<0\ \ \ \ \Rightarrow \ \  \OLo<\nu<1\,.
\end{equation}
Being $\aX>0$ we see that this choice is compatible with the
existence of the maximum at $\zm$, see Eq.(\ref{maximum}). In
contrast to scenario I, we find that to achieve stopping it is
not necessary that $\OLo<0$. This novel trait appears thanks to
the non-vanishing interaction between the two components of the
DE in the $\CC$XCDM, which  is gauged by the parameter $\nu$ --
as seen from the first two equations in (\ref{seteq}). Therefore,
we conclude that for quintessence-like cosmon we can have stopping
of the expansion not only for $\OLo<0$ but also for
$0<\OLo<\nu<1$.

\item V) $0<\aX<2$ and $\nu=1$. In this case we have the same eigenvalues
as in scenario III, but the situation is different. The
asymptotic node for the trajectories is
\begin{equation}\label{nu1node}
{\bf \Omega}^{*}_{V}(\nu)=\begin{pmatrix}
  1-\OLo \\
 \OLo-\OMo\,(1-\aX/\amr)\\
  1-\OMo\,(1-\aX/\amr)
\end{pmatrix}\,.
\end{equation}
Using the sum rule (\ref{sumrule0}) it is easy to see that for
$\aX=0$ this fixed point coincides with (\ref{alfa0de}) for
$\nu=1$, as it should. In contrast to scenario III, in this case
the stopping is in principle possible by requiring that the third
component of (\ref{nu1node}) becomes negative. This condition
yields
\begin{equation}\label{condV}
\aX<\frac{\OMo-1}{\OMo}\ \amr\,.
\end{equation}
Using $\amr=3$ for the matter-dominated epoch and the prior
$\OMo=0.3$, we find $\aX<-7$ (i.e. $\wX< -1.33$) to achieve
stopping. This would be no problem, were it not because the
nucleosynthesis constraint derived in Section
\ref{sect:nucleosynthesis} forbids that the product $|\nu\,\aX|$
is too large as in the present case. So this scenario is actually
ruled out by nucleosynthesis considerations.

\item VI) $0<\aX<2$ and $\nu>1$. Here one eigenvalue is positive
and the other negative ($\lu>0\,,\ld<0$) as in scenario II.
Therefore the point (\ref{nu0node}) is again a saddle point from
which all trajectories diverge as $\zeta\rightarrow\infty$. This
runaway, however, can be stopped provided $C_1<0$ in
(\ref{solve2}). Indeed, since the eigenvector ${\bf v_1}$ defines
a runaway direction the third component of (\ref{solve2}) will
eventually become negative, which is the condition for stopping.
In the absence of the nucleosynthesis constraint (\ref{eb}), the
condition $C_1<0$ would be a bit messy, but thanks to this
condition we have in good approximation
\begin{equation}\label{C1approx}
C_1=
\frac{1-\OLo}{1-\nu}-\frac{\OMo(\wm-\wX)}{\wm-\wX+\epsilon}\simeq
\frac{1-\OLo}{1-\nu}-\OMo\,,
\end{equation}
and
\begin{equation}\label{C2approx}
C_2= \frac{\OMo(\wm-\wX)}{\wm-\wX+\epsilon}\simeq \OMo\,.
\end{equation}
Thus, within this approximation, the stopping condition for this
case is
\begin{equation}\label{C1nucleoVI}
C_1<0\ \ \ \ \Longleftrightarrow \ \ \
\frac{1-\OLo}{1-\nu}<\OMo\,,
\end{equation}
where $\epsilon$ -- defined in (\ref{epsilon}) -- is assumed to be
small to satisfy the nucleosynthesis constraint. In the present
instance, since we assume $\nu>1$ we have to compensate it with
small enough $\aX$. The condition (\ref{C1nucleoVI}) is satisfied
by choosing
\begin{equation}\label{OLOM}
\OLo<1+(\nu-1)\,\OMo\,.
\end{equation}
This value could still perfectly lie in the preferred range of
values for $\OLo$ near the standard flat $\CC$CDM model, i.e.
$\OLo\simeq 0.7$\,\cite{Supernovae,WMAP03,LSS,WMAP3Y}. But of
course with a dramatic difference: with this same value we can
have turning point in the Universe evolution within the $\CC$XCDM
model. Combining the stopping condition (\ref{OLOM}) with the
condition (\ref{maximum}) for the existence of the maximum of
$r(z)$ under $\aX>0$, we see that we must have
$\OLo<1+(\nu-1)\,\OMo<\nu$, hence $\OMo<1$. Our usual prior for
this parameter ($\OMo=0.3$) automatically guarantees it.

\item VII) $-\delta<\aX<0$ and $\nu<1$. In this case we deal with
phantom-like cosmon. The eigenvalue $\lu$ is positive. Therefore
the point (\ref{nu0node}) is a saddle point, similar to scenario
VI. Again the stopping condition is (\ref{C1nucleoVI}). However,
here $\nu<1$ and thus we have
\begin{equation}\label{OLOM2}
\OLo>1-(1-\nu)\,\OMo\,.
\end{equation}
Using the prior $0<\OMo<1$ we find
\begin{equation}\label{OLOM3}
\OLo-\nu>1-\OMo-\nu\,(1-\OMo)>0.
\end{equation}
As $\aX<0$ the previous equation is compatible with
(\ref{maximum}), and so the simultaneous existence of the stopping
and the maximum is guaranteed.

\item VIII) $-\delta<\aX<0$ and $\nu>1$. Again we stick to
phantom-like cosmon here. The eigenvalue $\lu$ is nevertheless
negative. The trajectories are attracted to a node of the same
form as (\ref{nunode}). However, since $\nu>1$ in this case the
stopping condition takes the form
\begin{equation}\label{stopVIII}
\frac{\OLo-\nu}{1-\nu}<0\ \ \ \ \Rightarrow \ \  \OLo>\nu\,.
\end{equation}
On account of $\aX<0$ this choice is compatible with the existence
of the maximum at $\zm$, see Eq.(\ref{maximum}).

\end{itemize}
These are the main scenarios we wanted to stand out. From the
previous analysis it is clear that de Sitter space-time is not the
common attractor for the family of cosmological evolutions of the
$\CC$XCDM model, even in the absence of a turning point. Let us
notice the following interesting feature: the signs of $\OL(z)$
and $\OX(z)$ may change when we evolve from the past into the
future. This can be easily seen from (\ref{solve2}). For example,
in scenario VI (where $\nu>1$), and requiring stopping ($C_1<0$),
we have
\begin{eqnarray}\label{signVI}
&&\OL(z\rightarrow\infty)\simeq
\nu\,C_2\,(1+z)^{\amr}>0\,,\nonumber\\
&&\OL(z\rightarrow -1)\simeq
\nu\,C_1\,(1+z)^{\aX\,(1-\nu)}<0\,,\nonumber\\
&&\OX(z\rightarrow\infty)\simeq
-\frac{\nu\,C_2\,\amr}{\amr-\aX}(1+z)^{\amr}<0\,,\nonumber\\
&&\OX(z\rightarrow -1)\simeq
(1-\nu)\,C_1\,(1+z)^{\aX\,(1-\nu)}>0\,,
\end{eqnarray}
with $C_2\simeq \OMo>0$. We have also used $\amr>\aX$ -- see
Eq.\,(\ref{alfas}). The sign of the cosmological term goes from
positive in the past to negative in the future, and this is the
physical reason why the Universe halts in this case. On the
contrary the cosmon density in this same scenario is seen to be
negative in the past and positive in the future. At the turning
point $z_s$, where the expansion rate $H(z)$ vanishes, the sum
$\OD(z)=\OL(z)+\OX(z)$ takes on the negative value
\begin{equation}\label{sumLX}
\OD(z_s)=-\OMo\,(1+z_s)^3\,,
\end{equation}
and therefore it cancels against the matter
contribution\,\footnote{As warned in Sect.
\ref{sect:Lambdacosmon} the formal structure of (\ref{rDz})
should not lead us to think that $\rD=\rD(z)$ remains necessarily
positive in this model for the entire history of the Universe.
The effective EOS parameter $\we=\we(z)$ of the $\CC$XCDM model
(\ref{eEOS}) is actually singular at a point $z_0>z_s$ (i.e.
before stopping) where $\rD(z_0)=0$. Of course this is a spurious
singularity because all density functions remain finite at
$z=z_0$. After that point (i.e. for $z_s\leqslant z<z_0$) we have
$\OD(z)\lesssim0$, and at $z=z_s$ we have (\ref{sumLX}) exactly.
}. In contrast, in scenario VII (specifically for the situation
$\nu<0$) the sign changes of $\OL(z)$ and $\OX(z)$ from the past
to the future are just opposite to scenario VI, as can be easily
checked from the formulae above, so that here the cause for the
stopping is actually attributed to the negative energy of the
cosmon in the future (rather than to the effect of a negative
cosmological term). These are two complementary situations. In
both cases the cosmon field (or the effective entity it
represents) has negative energy either in the past or in the
future. It thus behaves in these cases as a phantom-like field
with negative energy, i.e. as Phantom Matter (cf. Fig.
\ref{fig1}). However, as in scenario II above, the effective
behavior of the $\CC$XCDM model remains quintessence-like because
$\aX\,\rX>0$.


\section{Crossing the $\we=-1$ divide?}
\label{sect:crossing}

The recent analyses \cite{Jassal1,Jassal2} of the observational
data imply the possibility of an interesting phenomenon known as
the crossing of the cosmological constant
boundary\,\cite{Phantom}. The essence of this phenomenon is that
the parameter of the dark energy equation of state, $\we=\we(z)$
at some positive redshift $z=z^{*}$ attains, and even trespasses,
the value $\we(z^{*})=-1$, i.e. the function $\we(z)$ crosses the
cosmological constant divide $\omega_{\CC}=-1$. Some analyses of
the cosmological data mildly favor the transition from the
quintessence-like to the phantom-like regime with the expansion
at a small positive redshift of order 1. Let us recall from the
analysis of Ref.\,\cite{SS1,SS2} that in models with variable
cosmological parameters we generally expect an effective crossing
of the cosmological constant barrier at some point near our time
(whether in the recent past or in the near future). This crossing
actually refers to the effective EOS parameter $\we=\we(z)$
associated to the model description in the effective ``DE
picture''\,\cite{SS2}, where the DE is self-conserved. This
picture is in contradistinction to the original cosmological
picture where $\CC$ is either exchanging energy with matter or is
varying in combination with Newton's gravitational constant
$G$\,\cite{SS2}. Of course in the original picture
$\omega_{\CC}=-1$, whether $\CC$ is strictly constant or
variable. The crossing, therefore, is visible only in the DE
picture, which is the one assumed in the common parametrizations
of the DE, and therefore the one used in the analysis and
interpretation of data.

What is the situation with the $\CC$XCDM model?  In the present
instance we do not necessarily expect such crossing. The reason is
that the $\CC$XCDM model is not a pure $\CC$ model (variable or
not) due to the presence of the cosmon; in fact, $\CC$ and $X$
form a non-barotropic DE fluid with $\we\neq -1$ from the very
beginning. For this reason the $\CC$XCDM model does not
necessarily follow the consequences of the theorem\,\cite{SS2}. It
may exhibit crossing of $\we=-1$, but it is not absolutely
necessary. As a matter of fact, in certain regions of the
parameter space there is no such crossing at all, but in other
regions there is indeed crossing. In the following we concentrate
again only on the flat space case. We shall prove next that
whenever there is crossing of $\we=-1$ within the $\CC$XCDM model
it can only be of a specific type in order to be compatible with
the existence of a turning point. Specifically, the crossing
should go from phantom-like behavior in the past into
quintessence-like behavior at present or in the future. It can
never be the other way around (if a turning point in the Universe
evolution must exist).

The proof goes as follows. From Eq.\,(\ref{eEOS}) we have $(1+\we)
\rD = (1+\wX)\,\rX$. The explicit solutions for the individual DE
densities $\rD(z)$ and $\rho_X(z)$ found in Sections
\ref{sect:solving1} and \ref{sect:solving2}  -- see e.g.
Eq.\,(\ref{solved1})-- reveal that these functions are nonsingular
for finite values of the scale factor, i.e. for the redshift
values in the entire range $-1 < z < \infty$. Therefore, in order
to have $\we(z) = -1$ at the transition point $z=z^{*}$, the
cosmon density must vanish at that point: $\rX(z^{*})=0$. The
explicit expression for $\OX(z)=\rX(z)/\rco$ (cf.
Eq.\,(\ref{solved1})) can be rearranged in the form
\begin{eqnarray}
&&\label{eq:fX} \frac{\OX(z)}{\OXo}=(1+b) (1+z)^{\aX\,(1-\nu)} -
b(1+z)^{\amr}\,,\nonumber\\
&& =
b\,(1+z)^{\amr}\,\left[\frac{1+b}{b}\,(1+z)^{\aX\,(1-\nu)-\amr}-1\right]\,,
\end{eqnarray}
where
\begin{equation}
\label{eq:b}
b=\frac{\nu\,(1+\omega_m)}{1+\omega_m-(1+\omega_X)(1-\nu)}\,\frac{\Omo}{\OXo}=
\frac{\nu\,(1+\omega_m)}{\wm-\wX+\epsilon}\,\frac{\Omo}{\OXo} \,.
\end{equation}
For $\nu=0$ it is patent that the expression (\ref{eq:fX}) shrinks
to (\ref{Oxnu0}) and it does never vanish for $z>-1$. In
contrast, for $\nu\neq 0$ the two terms in (\ref{eq:fX}) may
conspire to produce a zero. From the structure of $\OX(z)$ it is
clear that $\OX(z)$ can have at most {\em one} zero in the entire
interval $-1 < z < \infty$, namely at $z=z^{*}$ given by
\begin{equation}\label{cza}
1+z^{*}=\left\{\frac{b}{1+b}\right\}^{\frac{1}{\aX\,(1-\nu)-\amr}}\,.
\end{equation}
Consequently, the crossing of the $\we=-1$ line can happen at most
at one value in this redshift interval. Let us now concentrate on
the case when the DE evolves with the expansion of the universe
from the quintessence regime into the phantom regime at some
positive redshift $z^{*}$. In this case for $z > z^{*}$ we have
$\we> -1$ whereas for $z <z^{*}$ we have $\we < -1$ and the
present value of $\we$ is phantom-like, i.e. $\we(0) < -1$. Since
by assumption the crossing happened in the past at some $z^{*}>0$,
there can be no crossing in the future, since there is at most one
crossing. It follows that in the future $\we$ remains below $-1$,
and since the present DE density is positive in our model
($\rDo>0$), the function $\rD=\rD(z)$ can only grow to larger and
larger positive values with the expansion of the Universe -- see
Eq.\,(\ref{rDz}). Therefore the expansion rate $H$ given by
(\ref{FL}) will also keep growing forever, and hence no stopping
is possible. This is tantamount to say that $\we$ will exhibit no
singularities in the future, which is a necessary condition to
trigger stopping of the expansion. We conclude that this type of
cosmological constant boundary crossing is not compatible with the
condition of stopping of the expansion (\textit{q.e.d.})

On the other hand the alternate transition from the phantom-like
to the quintessence-like regime with the expansion of the Universe
is characterized by $\we(0) > -1$, for  $z^{*}>0$. This type of
crossing in the past ($z^{*}>0$) is perfectly compatible with the
stopping condition, and we can ascertain analytically under what
conditions it takes place. From Eq.\,(\ref{cza}) we see that in
principle the crossing can only take place under one of the two
following conditions:
\begin{equation}\label{options}
(i)\ \ b>0 \ \ {\rm and} \ \  \aX\,(1-\nu)-\amr<0,\ \  \ {\rm or}\
\ \ (ii)\ \ b<-1 \  \ {\rm and}  \ \  \aX\,(1-\nu)-\amr>0\,.
\end{equation}
However, option (ii) is impossible because the restriction
(\ref{wxrange}) on the range of $\wX$ and the nucleosynthesis
condition (\ref{eb}) imply together that
\begin{equation}\label{restriction}
\aX\,(1-\nu)-\amr=3\,\left(\wX-\wm-\epsilon\right)<0\,.
\end{equation}
Hence we are left with the condition $b>0$. From (\ref{eq:b}) we
finally conclude that the desired crossing in the past will take
place whenever $\OXo$ and $\nu$ have the same sign:
\begin{equation}\label{crossing}
 (\OXo>0,\ \ \nu>0)\ \ \ \ \ {\rm or}\ \ \ \ \  (\OXo<0,\ \ \nu<0)\,.
\end{equation}
Furthermore, on comparing the condition for the existence of the
crossing (\ref{cza}) with the stopping conditions listed in
Section \ref{sect:solving2} we see that they are compatible.
Explicit examples of this kind of crossing are illustrated in the
numerical analysis of Section \ref{sect:numerics}.

As stated above, some analyses of the cosmological data mildly
favor the transition from the quintessence-like to the
phantom-like regime, which is opposite to the kind of behavior we
find for the present model (if the existence of both a transition
point and a turning point is demanded). However, the crossing
phenomenon has never been established firmly to date, and the
more recent analyses\,\cite{Jassal2} cast some doubts on its
existence. Actually, the same kind of analyses leave open the
possibility that the crossing feature, if it is there at all, can
be of both types provided it is sufficiently mild; that is to
say, we can not exclude that in the recent past the Universe
moved gently (i.e. almost tangent to the $\we=-1$ line) from an
effective phantom regime into an effective quintessence regime.
In such case the average value of $\we(z)$ around $z=z^{*}$ would
stay essentially $\we=-1$ (therefore mimicking the standard
$\CC$CDM). Interestingly enough, the $\CC$XCDM model can perfectly
accommodate this possibility (see next section for explicit
numerical examples), but at the same time the model is also
perfectly comfortable with the absence of any crossing at all
while still maintaining a turning point in its evolution.



\section{Numerical analysis of the $\CC$XCDM model}
\label{sect:numerics}

\subsection{Parameter space} \label{sect:pspace}

In the previous sections we have solved analytically the $\CC$XCDM
cosmological model and we have performed a detailed analysis of
its analytical and qualitative properties. In the present section
we want to exemplify these properties by considering some
numerical examples. In the first place it is important to check
that the region of parameter space that fulfills all the necessary
conditions is a significant part of the full parameter space of
the $\CC$XCDM model. The latter is in principle defined by the
following $5$ parameters
\begin{equation}\label{parameterspace1}
(\OMo,\OLo,\OXo,\wX,\nu)\,.
\end{equation}
However, we will take a prior on $\OMo$ because the matter content
of the Universe can be accounted for from (galaxy) clusters data
alone\,\cite{LSS}, i.e. from LSS observations independent of the
value of the DE measured from, say, supernovae or CMB. The
approximate current value is \,\cite{LSS}: $\OMo=0.3$. Moreover,
since we want to restrict ourselves to the flat space case
($\OKo=0$), we find from the sum rule (\ref{sumrule0}) that
\begin{equation}\label{ODo}
\ODo=\OL^0+\OX^0=0.7\,.
\end{equation}
This choice of the cosmological parameters will be implicit in
\textit{all} our numerical analyses. We can then take e.g. $\OLo$
as an independent parameter. Thus in practice we are led to deal
with the basic three-dimensional parameter space
\begin{equation}\label{parameterspace2}
(\OLo,\wX,\nu)\,.
\end{equation}
It is on this smaller parameter space that we must impose the
necessary conditions to define the ``physical region'' of our
interest. These conditions are essentially three:
\begin{itemize}
\item i) The nucleosynthesis bound (cf Section
\ref{sect:nucleosynthesis})
\begin{equation}\label{Nbound}
|r_N|<10\%\,,
\end{equation}
i.e. the condition that the DE density is sufficiently small as
compared to the matter density at the nucleosynthesis time;

\item ii) The stopping of the expansion, i.e. the condition of reaching a
stopping point in the Universe evolution from which it can
subsequently reverse its motion (see
Section\,\ref{sect:solving2});

\item iii) The condition that the ratio of the DE energy density to the matter
density -- see Eq.\,(\ref{ratio}) -- is bounded and stays
relatively small throughout the entire history of the Universe,
say
\begin{equation}\label{ratiobound}
 \frac{r(z)}{r_0}<10\ \ \ \ (-1<z<+\infty)\,,
\end{equation}
$r_0$ being its present value (see Eq.\,(\ref{rz0}) and Section
\ref{sect:nucleosynthesis} in general).
\end{itemize}

\FIGURE[t]{ \centering
\resizebox{0.8\textwidth}{!}{\includegraphics{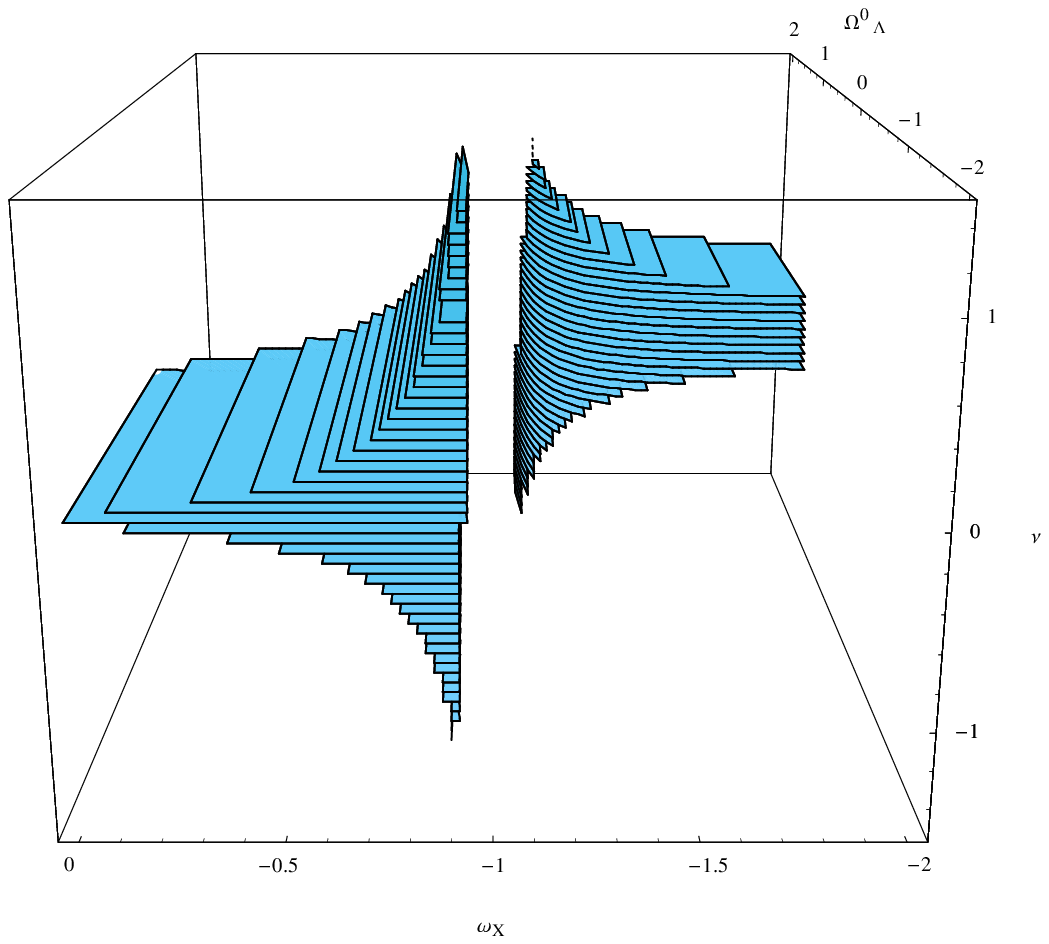}}
\caption{The $3D$-dimensional ``physical region'' of the $\CC$XCDM
model, consisting of the subset of points (\ref{parameterspace2})
satisfying the three requirements: i) Nucleosynthesis bound
(\ref{Nbound}), ii) Existence of a turning point in the Universe
evolution, and iii) Ratio of DE density to matter density,
$r=\rD/\rmr$, under the bound (\ref{ratiobound}) in the entire
Universe lifetime\,. }\label{fig2} }

\FIGURE[t]{
\mbox{\resizebox*{0.45\textwidth}{!}{\includegraphics{figure2a.eps}}\
\ \ \
   \resizebox*{0.45\textwidth}{!}{\includegraphics{figure2b.eps}}}\\

   \\
   \\
   \mbox{\resizebox*{0.5\textwidth}{!}{\includegraphics{figure2c.eps}}}
\caption{The projections of the $3D$-dimensional physical region
of Fig.\,\ref{fig2} onto the three planes: \textbf{(a)}
$\OLo=const.$, \textbf{(b)} $\wX=const.$ and \textbf{(c)}
$\nu=const.$  } \label{fig3}}

\FIGURE[t]{
\mbox{\resizebox*{0.45\textwidth}{!}{\includegraphics{figure3a.eps}}\
\ \ \
   \resizebox*{0.45\textwidth}{!}{\includegraphics{figure3b.eps}}}\\

   \\
   \\
   \mbox{\resizebox*{0.5\textwidth}{!}{\includegraphics{figure3c.eps}}}
\caption{Different slices of the $3D$ volume representing the
physical region, see Fig.\,\ref{fig2}. The dotted areas fulfill
the nucleosynthesis condition. The shaded regions satisfy both the
nucleosynthesis and stopping conditions. If we further require
the condition (\ref{ratiobound}) on the ratio $r=\rD/\rmr$, we
are left with the darkest shaded region. Slices shown in the
figure:\ \textbf{(a)} Slice $\Omega^0_{\Lambda}=0.75$;
\textbf{(b)} Slice $\omega_X=-1.85$; \textbf{(c)} Slice
$\nu=-\nu_0$, where $\nu_0$ is defined in
(\ref{nu0}).}\label{fig4} }

\FIGURE[t]{ \centering
\resizebox{0.8\textwidth}{!}{\includegraphics{figure4.eps}}
\caption{The Hubble function $H=H(z)$ for the values of the
parameters: $\wX=-1.85,\,\OLo=0.75,\nu=-\nu_0$. This situation
illustrates an example of Scenario VII (see Section
\ref{sect:solving2}). Condition (\ref{OLOM2}) is satisfied and
hence there is a stopping point in the future where $H$ vanishes
and the expansion of the Universe stops. } \label{fig5}}

The ``physical region'' is defined to be the subregion of the
parameter space $(\ref{parameterspace2})$ satisfying these three
conditions at the same time. It is illustrated in the $3D$-plot in
Fig.\,\ref{fig2}. In Fig.\,\ref{fig3} we can see the projection of
the full $3D$ physical region onto three orthogonal planes
$\OLo=const.$, $\wX=const.$ and $\nu=const.$  situated outside
this region. It is also convenient to show cross-sections of the
physical region, i.e. slices of the relevant volume for fixed
values of the other parameters, just to check that the physical
region is not just formed by a closed $2D$-surface, but rather by
a $3D$-volume contained in the space (\ref{parameterspace2}). In
Fig.\,\ref{fig4} we show three slices $\OLo=const.$, $\wX=const.$
and $\nu=const.$ for particular values of these parameters that
illustrate this feature. We note that in Fig.\,\ref{fig3} we have
allowed for $\wX$ to be slightly out in the range (\ref{wxrange}),
only for completeness. In fact the region that does not satisfy
condition (\ref{wxrange}) is seen to become narrower the larger
becomes $\wX$, and therefore it is rather restricted.

\subsection{Expansion rate and effective EOS} \label{sect:eEOS}

In Fig.\,\ref{fig5} we show the expansion rate $H=H(z)$ of the
$\CC$XCDM model as a function of the redshift for given values of
the parameters: $\wX=-1.85,\,\OLo=+0.75,\,\nu=-\nu_0$, where
$\nu_0$ is defined in (\ref{nu0}). This selection of parameters
corresponds to scenario VII in Section \ref{sect:solving2}. Since
Eq.\,(\ref{OLOM2}) is fulfilled, there must be a stopping point
ahead in time, which is indeed confirmed in Fig.\,\ref{fig5} where
we see that $H$ vanishes at some point in the future. As discussed
in Section\,\ref{sect:solving2}, the cause for the stopping in
this case is actually attributed to the negative energy of the
cosmon in the future (i.e. to the presence of ``Phantom Matter''
rather than to the effect of a negative cosmological term). The
presence of a stopping point implies the existence of a maximum
for the ratio between DE and matter densities, and this fact
provides a natural explanation for the coincidence problem. In the
present case the maximum exists because Eq.\,(\ref{OLOM3}) is
satisfied. Notice that the choice of parameters used in
Fig.\,\ref{fig5} has the attractive property that the value of
$\OL^0$ is very close to the current one in the standard $\CC$CDM
model. However, it should be remembered that in all numerical
examples discussed in this section we have fixed $\ODo$ as in
(\ref{ODo}). This means that in the particular numerical example
illustrated in Fig.\,\ref{fig5}, the value of the cosmon density
at present is rather small and negative: $\OXo=-0.05$.

\FIGURE[t]{
\mbox{\resizebox*{0.45\textwidth}{!}{\includegraphics{figure5a.eps}}\
\ \ \
   \resizebox*{0.45\textwidth}{!}{\includegraphics{figure5b.eps}}}\\

   \\
   \mbox{\resizebox*{0.45\textwidth}{!}{\includegraphics{figure5c.eps}}\
\ \ \
   \resizebox*{0.45\textwidth}{!}{\includegraphics{figure5d.eps}}}
\caption{The effective EOS parameter of the $\CC$XCDM model, $\we$
--defined in Eq.\,(\ref{eEOS}) --, as a function of the
cosmological redshift, for different scenarios: \textbf{(a)}
$\wX=-1.85,\,\OLo=0.75,\,\nu=-\nu_0$; \textbf{(b)} As in (a) but
for $\nu=0$; \textbf{(c)} As in (a), but for $\nu=+\nu_0$;
\textbf{(d)} $\wX=-0.93,\OLo=-2,\,\nu=+0.96$.
  } \label{fig6}}

Very important for the future program of cosmological
experiments\,\cite{SNAP,PLANCK,DES} is the issue of the equation
of state (EOS) analysis of the dark energy from the observational
data. In the standard description of the DE, it is assumed that
the DE consists of a self-conserved entity, or various entities of
the same kind which produce a conserved total DE. As a particular
case we have the standard $\CC$CDM model, where the DE is just the
constant $\CC$. But we can also have simple quintessence models,
with a single scalar field or multicomponent scalar
fields\,\cite{Copeland06,Ratra}. In the latter case one of the
fields can be a (standard) phantom (cf. Fig.\,\ref{fig1}) and this
has been proposed as a mechanism to explain the crossing of the
cosmological constant divide\,\cite{Phantom}. In practice the
experimental EOS
\begin{equation}\label{EOSexp}
\pD=\omega_{\rm exp}\,\rD\,,
\end{equation}
has to be determined by observations. This equation will be
confronted with the theoretical predictions of any given model.
For the theoretical analysis it is convenient to consider the
general notion of effective EOS of the DE\,\cite{{LinderEff}},
namely the EOS that describes dark energy as if it would be a
single self-conserved entity without interaction with matter . We
write it as
\begin{equation}\label{weff}
\pD=\we(z)\,\rD,
\end{equation}
where $\we=\we(z)$ is the theoretical EOS parameter or function in
the effective DE picture, which is the one that should be
compared with $\omega_{\rm exp}$ in (\ref{EOSexp}). In the
original model the DE could actually be a mixture of entities of
very different kind, as indicated in (\ref{mixture}). This is
indeed the case for the $\CC$XCDM model where we have a
non-barotropic mixture of cosmon and running cosmological
constant.  From the explicit solution of the model obtained in
Section \ref{sect:solving1} we can study the numerical behavior
of the effective EOS of the $\CC$XCDM model, see
Eq.\,(\ref{eEOS}).

Before displaying our EOS results, we recall that the most recent
WMAP data, in combination with large-scale structure and
supernovae data, lead to the following interval of values for the
EOS parameter\,\cite{WMAP3Y}:
\begin{equation}\label{weWMAP06}
\omega_{\rm exp}=-1.06^{+0.13}_{-0.08}\,.
\end{equation}
This result is quite stringent, and it does not depend on the
assumption that the Universe is flat. However, it \textit{does}
depend on the assumption that the EOS parameter does \textit{not}
evolve with time or redshift. In this sense it is not directly
applicable to the effective EOS of the $\CC$XCDM model. In
general the experimental EOS parameter $\omega_{\rm exp}$ need
not be a constant, and one expects it to be a function of the
redshift, $\omega_{\rm exp}=\omega_{\rm exp}(z)$. Since this
function is unknown, one usually assumes that it can be
parametrized as a polynomial of $z$ of given order. This is,
however, inconvenient to fit cosmological data at high redshifts;
say, data from the CMB. And for this reason other
parametrizations have been proposed and tested in the
literature\,\cite{Alam, Hannestad,Jassal1}. As an example,
consider the following one\,\cite{Jassal1},
\begin{equation}\label{weparam}
\we(z)=\omega_0+\omega'_0\,\frac{z}{(1+z)^p}\,, \ \ \ (p=1,2)\,,
\end{equation}
where $\omega'_0=(d\we/dz)_{z=0}$. The last parameter contemplates
the possibility of a residual evolution of the DE even if
$\omega_0=-1$. Notice that the asymptotic behavior of $\we(z)$ in
this parametrization is tamed: for $p=1$ one has
$\we(+\infty)=\omega_0+\omega'_0$, and for $p=2$,
$\we(+\infty)=\omega_0$. This kind of parametrizations can be
useful for a simplified treatment of the data. However, in general
one should not be misled by them. The underlying dark energy EOS
can be more complicated and may not adapt at all to a given
parametrization. In this respect it should be kept in mind that
there is also the possibility that the experimental EOS
(\ref{EOSexp}) is mimicked by other forms of DE that look as
dynamical scalar fields without being really so. This possibility
has been demonstrated in concrete examples in \cite{SS1} and
further generalized in \cite{SS2}. In particular, in the analysis
of \cite{SS1} it is shown that the ``effective EOS of a running
cosmological model'' can be analytically rather complicated; it
does not adapt to the form (\ref{weparam}) and nevertheless it
provides a good qualitative description of the experimentally
fitted EOS from the data\,\cite{Alam}. In the following we will
show, with specific numerical examples, that the effective EOS of
the $\CC$XCDM model may account for the observed behavior of the
DE in the redshift range relevant to supernovae experiments. In
Section \ref{sect:EOSasymptotic} we will show that even the
asymptotic behavior of the $\CC$XCDM model at very high redshifts
(relevant to the CMB analysis) completely eludes the class of
models covered by simple  parametrizations like (\ref{weparam}).
These parametrizations are, therefore, not model-independent, and
the fits to the dark energy EOS derived from them should be
handled with care and due limitation.

In Fig.\,\ref{fig6} we plot $\we=\we(z)$ for the $\CC$XCDM model
as a function of $z$ for some particular choices of the basic set
of parameters (\ref{parameterspace2}) and within the typical
accessible supernovae range\,\cite{SNAP}. On comparing these
results with the experimental fit (\ref{weWMAP06}) we should
stress again that the latter is only valid for a strictly constant
EOS parameter. Even so, scenarios (a)-(c) of Fig. \ref{fig6}
satisfy this bound already at $1\,\sigma$. Scenario (d) satisfies
it only at $\sim 2.5\sigma$, but here $\we$ varies substantially
with $z$.

All the plots in Fig. \ref{fig6} exhibit a discontinuity at a
certain point $z_0$ in the future, before reaching the stopping
point $z_s$. A magnified view of this discontinuity for the case
of Fig. \ref{fig6}a can be appreciated in Fig. \ref{fig7}. The
discontinuity occurs when $\we$ changes sign at a point $z_0$
where $\rD(z_0)=0$. At this point $\we(z_0)$ becomes undefined,
see Eq.\,(\ref{eEOS}). In a neighborhood
$z_0^{\pm}=z_0\pm\varepsilon$ of the point $z_0$ we have
$\we(z_0^{\pm})=\pm\infty$ for $\varepsilon\rightarrow 0$. The
singularity is not a real one, because all density functions
remain finite at that point, so its existence only reminds us
that the description in terms of the EOS parameter is inadequate
near that point.

\FIGURE[t]{
\centerline{\mbox{\resizebox*{0.5\textwidth}{!}{\includegraphics{figure5e.eps}}}
}

\caption{ Detail of the discontinuity of the EOS function
$\we=\we(z)$ at a point $z_0\lesssim -0.638$ in the future,
corresponding to case (a) of Fig. \ref{fig6}. At this point
$\rD(z_0)=0$.} \label{fig7} }

Looking at Fig. \ref{fig6} we can see diverse scenarios. For
example, in Fig. \ref{fig6}a we have a concrete example of the
scenario VII (with $\nu<0$) defined in Section
\ref{sect:solving2}, the same one used in Fig.\,\ref{fig5}, namely
$\wX=-1.85,\OLo=0.75,\nu=-\nu_0$. It corresponds to have $\OLo$
very near the standard $\CC$CDM value with the $\nu$ parameter
fixed at the canonical value (\ref{nu0}). At the same time the
cosmon field, $X$, is phantom-like and it behaves as Phantom
Matter in the future.  We can also appreciate a mild transition
of $\we$ from phantom to quintessence behaviour at some point
$z^{*}$ near past. This is the kind of possible transition we
have foreseen analytically for the $\CC$XCDM model in Section
\ref{sect:crossing}. We can check that the crossing condition (i)
in Eq.\,(\ref{options}) is satisfied. If the transition is
sufficiently mild and the average of $\we$ around the transition
point $z^{*}$ is close to $-1$  (as it is indeed the case here)
it cannot be excluded by the present data. Notice that the
phantom regime starts to be significant only for $z>1.5$ in this
case. At the moment we do not have enough statistics on this
region, and it is not obvious that we will ever have, at least
using supernovae data alone. Recall that SNAP\,\cite{SNAP} is
scheduled to possibly reach up to $z=1.7$, with a very meager
statistics foreseen around this high redhshift end. In this
sense, a mild phantom behavior around that distant regions will
be difficult to exclude; in particular, a kind of smooth
transition of the type Phantom $\rightarrow$ Quintessence (for
increasing $z$) will always be a possibility to keep in mind.
Notice that in all cases shown in Fig. \ref{fig6} the value of
$\we(z)$ at $z=0$ is very close to $-1$, namely $\we(z)=-0.94$ in
the first three cases, and $\we(z)=-0.73$ in the last one. As we
have said, these values are sufficiently close to $-1$ to be
acceptable in the light of the recent data and the priors used to
fit it.

In Fig.\,\ref{fig6}b we display a situation corresponding  to
$\nu=0$. This is an example of scenario II in Section
\ref{sect:solving2}. In this case the cosmological term $\CC$
remains strictly constant and the stopping of the Universe is due
to the super-deceleration produced by the cosmon (mainly in the
future time stretch). Again $\OXo<0$, but here the cosmon behaves
as Phantom Matter for the entire span of the Universe lifetime.
For $\nu=0$, we can see that in the past $\we$ approaches fast to
$-1$ asymptotically. The reason for this can be seen e.g. in
Eq.\,(\ref{signVI}) -- and more generally from (\ref{eq:fX}) --
which shows that in the asymptotic past the cosmon density
$\OX(z)$ is proportional to $\nu$. Figures \,\ref{fig6}c,d display
other interesting situations. They constitute examples of
scenarios VII and IV respectively (both with $\nu>0$). In
particular, in Fig.\,\ref{fig6}c the EOS parameter approaches
$\we=-1$ for a long stretch of the accessible redshift range by
SNAP, and therefore it mimics a constant cosmological term. This
case is different from the one in Fig.\,\ref{fig6}b, where $\CC$
is strictly constant but the DE is not; here the $\CC$XCDM model
behaves as the standard $\CC$CDM model for sufficiently large
redshift. Finally, in \ref{fig6}d we have a situation where the
cosmon is quintessence-like ($\wX>-1$) and nevertheless the
effective behavior of the total DE displays a transition from
phantom to quintessence regime. Again we can check that the
crossing condition (i) in Eq.\,(\ref{options}) is satisfied.

 \FIGURE[t]{
\mbox{\resizebox*{0.45\textwidth}{!}{\includegraphics{figure6a.eps}}\
\ \ \
   \resizebox*{0.45\textwidth}{!}{\includegraphics{figure6b.eps}}}
\caption{The total and individual DE densities, $\OD=\OD(z)$,
$\OL=\OL(z)$ and $\OX=\OX(z)$ for two different situations where
the cosmon barotropic index is of phantom-type ($\wX<-1$) and of
quintessence-type ($\wX\gtrsim-1$) respectively:\ \textbf{(a)}
$\wX=-1.85,\,\OLo=0.75,\,\nu=-\nu_0$;
 \textbf{(b)}  $\wX=-0.93,\,\OLo=-2,\,\nu=0.96$.
The two examples correspond to cases (a) and (d) of
Fig.\,\ref{fig6}.}.

\label{fig8}}

\subsection{Evolution of the DE density and its components} \label{sect:DEdensities}

The behavior of the effective EOS parameter $\we$ is related to
that of the DE densities in Eq.\,(\ref{eEOS}). The interaction of
the cosmon and the cosmological term produces a curious effect;
namely, the evolution of the cosmon density, $\OX$, does not
follow the expectations that we would usually have for a fluid $X$
carrying a value $\wX>-1$ or $\wX<-1$ of its barotropic index. For
a better understanding of the situation we have plotted in
Fig.\,\ref{fig8}a,b the total density $\OD(z)$, as well as the two
individual components $\OL(z)$ and $\OX(z)$, for two particular
cases: one with the same parameters as in Fig.\,\ref{fig6}a (in
which the cosmon has $\wX<-1$ as if it were a standard phantom
field), and the other for the same parameters as in
Fig.\,\ref{fig6}d (where the cosmon has $\wX>-1$ as if it were a
conventional quintessence field). We can appreciate from these
figures the sign changes of the various DE densities with the
Universe evolution. In particular, in Fig.\,\ref{fig8}a the cosmon
density appears to be $\OX>0$ in the intermediate past but it
starts behaving as Phantom Matter ($\OX<0$, cf. Fig.\,\ref{fig1})
near our recent past and also at present, and this trend is
enhanced in the future. In contrast, for the case of
Fig.\,\ref{fig8}b the $X$ density changes fast from $\OX<0$ to
$\OX>0$ at some point in the past (and stays so all the way into
the future). Now, due to its interaction with the variable $\CC$,
the quintessence-type cosmon ($\wX\gtrsim -1$) of
Fig.\,\ref{fig8}b does not decay with the expansion; it actually
grows fast as if it were a true phantom! The overall result is
that, thanks to the concomitant depletion of the $\CC$ density
into the negative energy region (very evident in the figure),
there is a net deceleration of the expansion and a final arrival
at the stopping point. Notice that, in contrast to the hectic
evolution of the two component $\OL(z)$ and $\OX(z)$, the total
DE density $\OD(z)$ evolves quite slowly. A different behavior is
observed for the phantom-type cosmon ($\wX< -1$) in
Fig.\,\ref{fig8}a. In this case, rather than increasing rapidly
with the expansion, the cosmon density actually decreases slowly
(it actually behaves for a while as standard quintessence) and
then at some future point near our time it ``transmutes'' into
Phantom Matter and causes a fast deceleration of the expansion,
with an eventual stopping of the Universe. In this case, the
total DE density $\OD(z)$ again evolves very mildly. The
behaviors in Fig.\,\ref{fig8}a,b are an immediate consequence of
Eq.\,(\ref{wXrX}), which implies that the $\CC$XCDM model
effectively behaves as quintessence if and only if $\aX$
\textit{and} $\OX$ have the same signs, as it is indeed the case
here.

Recall that in all cases considered in our numerical analysis the
total DE density is normalized to (\ref{ODo}) at $z=0$, as can be
checked in the examples presented in Fig.\,\ref{fig8}. In this
figure it is remarkable the stability of $\OD(z)$ in the entire
range relevant to SNAP experiments\,\cite{SNAP} both for small
$\nu$ (Fig.\,\ref{fig8}a) and large $\nu$ (Fig.\,\ref{fig8}b).
This is due to the nucleosynthesis constraint and to the small
values of the cosmon parameters $\aX$ and/or $\OXo$ in these
examples. To see this, let us first expand the total DE density
(\ref{ODz}) in the small parameter $\epsilon$ in first order:
\begin{eqnarray}\label{ODzsmall1}
\OD(z)
&&\simeq\frac{\OLo-\nu}{1-\nu}-\frac{\epsilon\,\Omo}{\wm-\wX}\,(1+z)^{\amr}
+\left\{\frac{1-\OLo}{1-\nu}-\Omo\right.\nonumber\\
&&\left.+\frac{\epsilon\,\Omo}{\wm-\wX}
-3\,\epsilon\,\left[\,\frac{1-\OLo}{1-\nu}-\Omo\right]\,\ln(1+z)\right\}
\,(1+z)^{\aX}\,. \end{eqnarray} Next, expanding this equation
linearly in $z$, the $\nu$-dependence cancels and we find a very
simple result:
\begin{equation}\label{ODzsmall2}
\OD(z)\simeq\ODo+\aX\,\OXo\,z\,.
\end{equation}
This expression coincides with the expansion, up to the linear
term in $z$, of the function (\ref{ODznu0}). The latter is the DE
density for the $\nu=0$ case -- see scenarios I and II of
Section\,\ref{sect:solving2}. Therefore, we arrive at the
following remarkable result: due to the nucleosynthesis
constraint, the DE density of the $\CC$XCDM model behaves
\textit{always} (i.e. for any $\nu$) as in the $\nu=0$ case in
the region of small $z<1$ (up to terms of order $\epsilon^2$). As
a consequence the evolution of $\OD$ with the redshift will be
mild in this region provided $\aX$ is small or $\OXo$ is small, or
both. Since $\nu$ has dropped in the linear approximation
(\ref{ODzsmall2}), this result is independent of $\nu$. Therefore,
$\nu$ can be large (meaning $|\nu|\sim 1$) and still have a
sufficiently ``flat'' evolution of $\OD(z)$ for $z\lesssim
1$\,\footnote{This situation is in contradistinction to the RG
model presented in Ref.\,\cite{RGTypeIa}, where the departure of
the DE density ($\rL$ density in that case) with respect to its
constant value in the standard $\CC$CDM model depends only on
$\nu$. In that model, we must have $|\nu|\ll 1$ if the DE
evolution is to be mild.}. We can appreciate particularly well
this feature in the situation depicted in Fig.\,\ref{fig8}a,
where the cosmon energy density is very low (in absolute value)
both in the intermediate past and also near the present. Thus
both the total DE density $\OD(z)$ and the cosmological term
density, $\OL(z)$, remain virtually constant for a long period in
our past and in our future - see the plateau $\OD\simeq const.$
in Fig.\,\ref{fig8}a. Although in this example the cosmon
density, $\OX$, remains very small in our past and around our
present ( $|\OX|\ll\OL$), in the future $\OX$ dives deeply into
the Phantom Matter region and eventually controls the fate of the
Universe's evolution. So the cosmon density, which is negligible
for the redshift interval relevant to SNAP measurements, does
become important in the future and is finally responsible for the
stopping of the Universe. Similarly, in Fig.\,\ref{fig8}b we have
a large value of $\nu$ around $1$ and $\OD$ stays also rather
flat. In this case the flat behavior is due to $\aX\ll 1$. In
both situations represented in Fig.\,\ref{fig8} the $\CC$XCDM
model mimics to a large extent the standard $\CC$CDM model.

\FIGURE[t]{
\centerline{\mbox{\resizebox*{0.8\textwidth}{!}{\includegraphics{ratcomc.eps}}}
} \caption{ Comparison of the effective EOS parameter of the
$\CC$XCDM model, $\we$, for fixed values of $\wX$ and $\OLo$
($\wX=-1.85$, $\OLo=0.75$, as in Fig.\,\ref{fig6}a) and different
values of $\nu$ in units of $\nu_0$, Eq.\,(\ref{nu0}). All curves
give $\we(0)=-0.94$ at the present time. } \label{fig9}}

Recall from Section \ref{sect:solving1} that in the limit of
vanishingly small $\nu$ \textit{and} $\OXo$ the $\CC$XCDM model
boils down exactly to the standard $\CC$CDM model. The result
proven above, however, tells us something new: namely, if $z$ is
not too large ($z\lesssim 1$, still within most of the SNAP
range), the coincidence between the two models can still be
maintained under appropriate circumstances (small $\aX$ and/or
$\OXo$), even if $\nu\neq 0$, by virtue of the nucleosynthesis
constraint on the parameter $\epsilon$. Therefore, the following
question is now in order. Under these circumstances, how can one
possibly differentiate between the two models in a practical way
(e.g. using the future SNAP experiments)? The answer is
straightforwardly simple: by looking closely at the EOS behavior
which, contrary to the DE density, can be very sensitive to the
value of $\nu$! In fact, the size and sign of $\nu$ can play a
key role to modulate the quintessence/phantom behavior
effectively developed by the $\CC$XCDM model at moderate values
of $z$ accessible to experiment. The EOS behavior has been
displayed in Fig.\,\ref{fig6}, but the $\nu$-sensitivity of this
behavior can be much better appreciated in the comparative set of
EOS curves shown in Fig.\,\ref{fig9}, which have been performed
for various values of $\nu$ (with both signs) and for fixed
values of the other parameters. For a cosmon of the ``$\wX<-1$
type'', we can see that a positive value of $\nu$ helps to lift
the EOS curve into the quintessence region and, therefore, it
helps to resemble more to some recent EOS analyses of the data,
which seem to favor the existence of a mild quintessence-regime
in our recent past \,\cite{Alam,Jassal1}. For $\nu<0$ (and small),
instead, the curve gently crosses the cosmological constant
divide and plunges slowly into the phantom regime at higher and
higher $z$. For a ``$\wX>-1$ type'' cosmon, instead, these
features are opposite (cf. Fig.\,\ref{fig6}d). Finally, for
$\nu=0$ the curve stabilizes very rapidly (at increasingly high
redshifts) along the flat line $\we=-1$ and the $\CC$XCDM model
looks in this case essentially as the standard $\CC$CDM both from
the point of view of EOS behavior and constancy of the DE density.
This is because $\aX<0$ in Fig.\,\ref{fig9}, meaning that
$\OD(z)\rightarrow\OL^0$ for $z$ sufficiently deep in the past
(see next section). But even in this case the effective EOS starts
changing softly at some point near our past, and then steadily
fast in the future, until it breaks into a singular (but
spurious) behavior at a point $z_0$ where $\rD(z_0)=0$ and
$\we(z_0)$ becomes undefined.

\subsection{Asymptotic regime of the EOS in the past} \label{sect:EOSasymptotic}

Let us study the behavior of the effective EOS in the asymptotic
past, $z\gg 1$. This will help to substantiate the behavior of the
numerical examples considered above and reveal additional
features. From Eq.\,(\ref{eq:fX}) we find
\begin{equation}
\OX(z\gg 1)=\left\{
\begin{array}{ll}
-b\,\OXo\,(1+z)^{\amr},
 & \, \, \, \, \, \, {\text{for}}\, \, \nu\neq 0 \\
 & \\
\hspace{1cm} 0 & \, \, \, \, \, \, {\text{for}}\, \, \nu= 0,\ \
\aX<0\\
& \\
\hspace{0.5cm}\OXo\,(1+z)^{\aX} & \, \, \, \, \, \,
{\text{for}}\, \, \nu= 0,\ \ \aX>0\,,
\end{array}
\right. \label{OXzinf}
\end{equation}
where $b$ is given by (\ref{eq:b}). Consider next the exact
expression for $\OD(z)$, given in (\ref{ODz}). The asymptotic
behavior of this expression in the past is
\begin{equation}
\OD(z\gg 1)=\left\{
\begin{array}{ll}
-\frac{\epsilon}{\wm-\wX+\epsilon}\,\Omo\,(1+z)^{\amr},
 & \, \, \, \, \, \, {\text{for}}\, \, \nu\neq 0 \\
 & \\
\hspace{1cm}\OLo & \, \, \, \, \, \, {\text{for}}\, \, \nu= 0,\ \
\aX<0\\
& \\
\hspace{0.5cm}\OXo\,(1+z)^{\aX} & \, \, \, \, \, \,
{\text{for}}\, \, \nu= 0,\ \ \aX>0\,.
\end{array}
\right. \label{ODzinf}
\end{equation}
For completeness let us also quote
\begin{equation}
\OL(z\gg 1)=\left\{
\begin{array}{ll}
\frac{\nu\,(\wm-\wX)}{\wm-\wX+\epsilon}\,\Omo\,(1+z)^{\amr}\simeq\nu\,\Omo\,(1+z)^{\amr}\,,
 & \, \, \, \, \, \, {\text{for}}\, \, \nu\neq 0 \\
 & \\
\hspace{1cm}\OLo & \, \, \, \, \, \, {\text{for}}\, \, \nu= 0\,.
\end{array}
\right. \label{OLzinf}
\end{equation}
We can indeed check out the predicted trends for the various DE
densities on inspection of Fig.\,\ref{fig8}. Substituting
equations (\ref{OXzinf}) and (\ref{ODzinf}) into (\ref{eEOS}),
and using (\ref{eq:b}), we find the  corresponding asymptotic
behavior of the effective EOS parameter in the past:
\begin{equation}
\we(z\gg 1)= -1+(1+\wX)\,\frac{\OX(z\gg 1)}{\OD(z\gg 1)}=\left\{
\begin{array}{ll}
\wm
 & \, \, \, \, \, \, {\text{for}}\, \, \nu\neq 0 \\
 & \\
\hspace{0.0cm}-1 & \, \, \, \, \, \, {\text{for}}\, \, \nu= 0,\ \
\aX<0\\
& \\
\hspace{0.0cm}\wX & \, \, \, \, \, \, {\text{for}}\, \, \nu= 0,\ \
\aX>0\,.
\end{array}
\right. \label{wezinf}
\end{equation}
On the one hand the asymptotic results for $\we$ in the case
$\nu=0$ were easily foreseeable from the structure of
Eq.\,(\ref{eEOS}), because in this case (that is, in the absence
of interaction between the cosmon and $\CC$) the growing and
decreasing properties of the cosmon density strictly follow its
canonical quintessence or phantom character in each case. In
other words, for $\nu=0$ a cosmon carrying a  phantom-type or
quintessence-type $\wX$-label does properly behave as a standard
phantom or quintessence, respectively. (Not so, however, when
$\nu\neq 0$, as we have seen!) On the other hand the asymptotic
result for $\we$ in the case $\nu\neq 0$ came a bit of a
surprise. It implies that at very high redshift the effective EOS
of the DE in the $\CC$XCDM model coincides with that of
matter-radiation. This does not cause any harm to
nucleosynthesis, for as shown by Eq.\,(\ref{ODzinf}) the
asymptotic DE density remains bounded by the constraint
(\ref{nb1}). This prediction of the $\CC$XCDM opens the
possibility that the CMB data could actually be blind to the EOS
analysis of the DE. Indeed, as we have mentioned above the CMB
fits usually assume parametrizations of the form (\ref{weparam})
and the like. In general one replaces Eq.\,(\ref{rDz}) into the
expansion rate (\ref{FL2}) to get (for $\OKo=0$)
\begin{equation}\label{FL2CMB}
H^2(z)=H_0^2\,\left[\Omo\,(1+z)^{\amr}+\,\ODo\,\exp\left\{3\,\int_0^z\,dz'
\frac{1+\we(z')}{1+z'}\right\}\right]\,.
\end{equation}
Whatever it be the ``well-behaved'' parametrization assumed for
$\we$ at high redshift, if it leads to powers of $1+z$ in the
function $H=H(z)$, the fact that $z\simeq 1100$ for the CMB
analysis will unavoidably provide a fit of $\we$ very close to
$\,-1$. For example, in the case of Eq.\,(\ref{weparam}) for
$p=1$, we find
\begin{eqnarray} \label{Hzzzquint} H^2(z)=
H_0^2\,\left\{\Omo\,(1+z)^{\amr}
+\ODo\,(1+z)^{3(1+\omega_0+\omega'_0)}\,\exp\left[-3\,\omega'_0\,\frac{z}{1+z}\right]
\right\} \,.
\end{eqnarray}
This equation reduces to the standard $\CC$CDM one for
$\omega_0=-1,\omega'_0=0$, as expected. At the same time it is
clear that fitting data at very high redshift (e.g. CMB data) with
(\ref{Hzzzquint}) will naturally convey $\we\simeq -1$ -- if one
wants to match at the same time the LSS and supernovae data both
suggesting that $\ODo$ is of the same order as $\OMo$. Therefore,
finding $\we\simeq -1$ from (\ref{Hzzzquint}) can be an intrinsic
bias or artifact of the parametrization itself.

In contrast, if the underlying theory of the DE would exhibit a
behavior of the $\CC$XCDM type (\ref{wezinf}), the CMB data could
not be sensitive to the EOS parameter in its conventional form
and the observations should be interpreted in a different way.
Namely, the canonical result $\we\simeq -1$ could only be derived
for ``small'' or intermediate redshift of the order of the distant
Type Ia supernovae, where the $\CC$XCDM model can resemble to a
large extent the standard $\CC$CDM. In the new scenario the result
(\ref{wezinf}) for $\nu\neq 0$ could still be detected by
observing a ``mass renormalization'' effect of the cosmological
parameter $\Omo$ when comparing the two sets of results for small
and high redshifts, i.e. when comparing supernovae data fits
with  CMB fits. Indeed, substituting (\ref{ODzinf}) in
(\ref{FL2}) one gets, at very high redshift (neglecting the
curvature term) \,\footnote{We note from the exact formula
(\ref{ODz}) that the total DE density in the $\CC$XCDM does not
scale in general as matter-radiation  $\sim (1+z)^{\amr}$, because
of the additional term $\sim (1+z)^{3\,(1+\wX-\epsilon)}$ which
plays an important role for $z\sim 1$. It is only at very high
redshift that the aforementioned scaling takes place. }:
\begin{equation}\label{renormalization}
H^2(z)\simeq H_0^2\,\hat{\Omega}_m^0\,(1+z)^{\amr}
\end{equation}
where the ``renormalized'' cosmological mass parameter is:
\begin{equation}\label{renormalized}
\hat{\Omega}_m^0=
\Omo\,\left(1-\frac{\epsilon}{\wm-\wX+\epsilon}\right)\,.
\end{equation}
The relative difference $|\hat{\Omega}_m^0-\Omo|/\Omo$ between
the two determinations of $\Omo$ is just given by the
nucleosynthesis constraint on the parameter $r=\rD/\rmr$, see
Eq.\,(\ref{nb1}). Thus it could lead to a measurable $\sim10\%$
effect that would provide a distinctive signature of the model.
There is nothing odd in the DE behaving this way; the DE density
(\ref{ODzinf}) is still there all the time even though the
conventional EOS analysis could miss it. In this case one must
resort to a different interpretation which may lead to its
determination.

Let us also note that since $\wm\geqslant0$, then in those
($\nu\neq 0$) cases where there is a crossing point  $z^{*}$ from
phantom into quintessence regime there must necessarily be a
point in the past where $\rD$ becomes zero and changes sign. This
produces another spurious singularity in the function $\we$.
(Recall from Section \ref{sect:crossing} that we cannot have more
than one crossing point $z^{*}$.) In this case the evolution of
$\we$ from the asymptotic regime $\we (z\gg 1)\simeq \wm>0$ up to
the singularity, is such that $\we$ stays above $-1$. After the
fake singularity is crossed towards $z=z^{*}$, the value of $\we$
emerging from it is smaller than $-1$, and remains so until
reaching the crossing point $z^{*}$ where again $\we$ becomes
larger than $-1$.

For experiments like SNAP\,\cite{SNAP} the relevant redshift
range (a window of small and moderate redshifts up to $z=1.7$) is
far away from the asymptotic regime. Here the rich panoply of
behaviors exhibited by the effective EOS of the $\CC$XCDM model
is very evident in e.g. Figs.\,\ref{fig6} and \ref{fig9}. Some of
these traits should in principle be observable in these
experiments and should be the most conspicuous ones. However, as
repeatedly emphasized, the effective EOS behavior of the
$\CC$XCDM model in that redshift range can approximate to a great
extent that of the $\CC$CDM model -- either because it is very
close to $\we=-1$ or because there is a gentle transition (from
phantom-like behavior into quintessence one) while keeping an
average value $\we=-1$.

\FIGURE[t]{
\mbox{\resizebox*{0.45\textwidth}{!}{\includegraphics{figure7a.eps}}\
\ \ \
   \resizebox*{0.45\textwidth}{!}{\includegraphics{figure7b.eps}}}\\

   \\
   \mbox{\resizebox*{0.45\textwidth}{!}{\includegraphics{figure7c.eps}}\
\ \ \
   \resizebox*{0.45\textwidth}{!}{\includegraphics{figure7d.eps}}}
\caption{The ratio $r=\rD/\rmr$, Eq.\,(\ref{ratio}), as a function
of the redshift (in units of the current ratio, $r_0$), for the
following values of the parameters: \textbf{(a)}
$\wX=-1.85,\OLo=0.75, \nu=-\nu_0$;\textbf{(b)} As before, but for
$\nu=0$; \textbf{(c)} The same, but for $\nu=+\nu_0$; \textbf{(d)}
$\wX=-0.93,\OLo=-2,\nu=0.96$.} \label{fig10} }

\subsection{Evolution of the ratio $r=\rD/\rmr$ and the cosmological
coincidence problem} \label{sect:revolution}

In Fig.\,\ref{fig10} we display the ratio $r$, Eq.\,(\ref{ratio}),
in units of the current ratio ($r_0\simeq 7/3$) as a function of
the cosmological redshift variable, for various values of the
parameters. All the plots in Figs.\,\ref{fig10} and \ref{fig11}
exhibit a maximum below 10. However, let us clarify that there are
regions of parameter space where the maxima can be much higher,
but in any case finite. Having a model where the ratio
$r=\rD/\rmr$ is bounded for the whole Universe lifetime provides a
natural explanation for the cosmological coincidence problem. One
may view the property $r_0\sim 1$ as a simple corroboration of the
fact that we live at a cosmic time near the transition redshift
$z_t$ where the Universe changed from decelerated expansion into
an accelerated one. But this observation leaves the coincidence
problem in the same place. For instance, in the standard $\CC$CDM
model, once the redshift $z_t$ is trespassed by the evolution the
ratio $r(z)$ starts to grow without end. So it is a kind of
puzzling ``coincidence'' that we just happen to live so near the
transition point $z_t$. This assertion should not be understood on
anthropic grounds. When we say that it is a coincidence to ``live
near the transition point $z_t$'' we mean that the probability to
find a lapse of the Universe's lifetime where $r(z)\simeq 1$
should be essentially zero for a Universe (like the standard
$\CC$CDM) where this ratio is larger than any given number if we
await enough time. In the $\CC$XCDM model, instead, the situation
is quite different. If the Universe would still be populated by
some form of intelligent life in the remote future, say $5-10$
billion years from now, and these beings would be curious enough
to perform a new measurement of $r(z)$ at their time, they might
learn that $r(z)$ is no longer $r_0$, as it is now, but maybe
something like $2\,r_0$ or $5\,r_0$, or even back to $r_0$ at
some later time (where the Universe would in this case be
decelerating rather than accelerating). There would be no
correlation between the value $r(z)\sim 1$ and the state of
acceleration of the Universe, and they would not be compelled to
conclude that they are living in a very peculiar spell of time.
We can confirm this picture by looking at some numerical examples
in Fig.\,\ref{fig10} and \ref{fig12}. In particular, in
Fig.\,\ref{fig12} we demonstrate the dependence of the maximum on
the value of $\nu$ for fixed values of the other parameters.

\FIGURE[t]{
\mbox{\resizebox*{0.7\textwidth}{!}{\includegraphics{ratcomb.eps}}}
\caption{Comparison of the maximum of the ratio $r=\rD/\rmr$ for
$\wX=-1.85$, $\OLo=0.75$ (as in Fig.\,\ref{fig10}a) and different
values of $\nu$.} \label{fig12} }

For a more physical grasp of the situation, we have also solved
numerically the $\CC$XCDM model in terms of the cosmic time
variable $t$ (cf. Fig.\,\ref{fig11}). In this case it is not
possible to provide analytic expressions such as (\ref{solved1})
or (\ref{solve2}) in terms of the $z$ and $\zeta$ variables
respectively. We limit ourselves to provide the numerical result
for the expansion rate $H=H(t)$ (Fig.\,\ref{fig11}a)  and the
ratio $r=r(t)$ (Fig.\,\ref{fig11}b) in units of the present ratio
$r_0$ for the cases (a),(b) and (c) of Fig.\,\ref{fig10}. We can
see that in terms of the cosmic time the distribution of the
relevant events in the future (such as the reaching of the maximum
and the arrival at the turning point) is not as sharp and abrupt
as it was in terms of the redshift variable. Indeed, one can see
that in terms of the cosmic time the maximum takes place far away
in the future, typically at around $2$ Hubble times ($H_0^{-1}$)
since the beginning of the expansion, i.e. roughly one Hubble
time after our time, where $H_0^{-1}\simeq 14\,Gyr$ for
$h_0=0.7$. The stopping point can typically lie at around $t_s\sim
(2.5-3)\,H_0^{-1}$. To avoid cluttering, scenario (d) of
Fig.\,\ref{fig10} is not included in Fig.\,\ref{fig11}, but we
have checked that it amounts to a lower maximum ($r<4$), and the
stopping point lies at $t_s\gtrsim 3.5\,H_0^{-1}$.

From our detailed analysis of the ratio $r=\rD/\rmr$ we may assert
that the $\CC$XCDM model provides a possible solution (and in any
case highly alleviates) the cosmological coincidence problem. At
the very least it offers an alternative explanation to some other
interesting proposals discussed in the
literature\,\cite{InteractingQ,OscillatingQ,Scherrer,Carneiro05}.
For example, in phantom dominated universes one may compute the
fraction of the lifetime of the Universe where the ratio
(\ref{rz}) stays within given bounds before the
doomsday\,\cite{Scherrer}. Recall that these universes are
threatened and eventually killed by a final Big Rip, so that the
idea is to compute the time interval between the start of
acceleration and the time when the Universe is dismantled to its
ultimate fundamental components, and compare it with the total
lifetime. If the fraction is not very small, one may feel
satisfied with this explanation. However, we can see that a
purely phantom Universe is just a very particular case of
$\CC$XCDM Universe, namely it corresponds to $\CC_0=\nu=0$ and
$\aX<0,\,\OXo>0$. In a general $\CC$XCDM Universe the range of
possibilities is much larger, e.g. there can be a turning point
rather than a Big Rip, the ratio (\ref{rz}) can stay bounded
during the entire span of the Universe lifetime, and finally the
matter is not destroyed.

Recall from the analytical discussion in Sections
\ref{sect:nucleosynthesis} and \ref{sect:solving2} that we have
three critical points in the future: $z_s$ (stopping/turning
point), $z_0$ (point where the total DE density vanishes and
$\we$ becomes singular), and $z_{\rm max}$ (point in the future
where the ratio $r(z)$ attains its maximum). They are ordered as
follows:
\begin{equation}\label{zzz}
z_s<z_0<z_{\rm max}\,.
\end{equation}
In correspondence we have the following values of the ratio
(\ref{ratio}):
\begin{equation}\label{rrr1}
r(z_s)=-1<r(z_0)=0<r(z_{\rm max})\lesssim 10\,.
\end{equation}
At the same time we may compare the evolution undergone by $r(z)$
from the nucleosynthesis time through the present time till the
maximum. We have exemplified typical situations in which
\begin{equation}\label{rrr2}
|r_N|\simeq 0.1<r_0\simeq 1<r(z_{\rm max})\lesssim 10\,.
\end{equation}
Namely, the ratio of the DE density to the matter density
undergoes a slow evolution within a few orders of magnitude. In
this context it does not look bizarre that its present value is
of order one. {It should, however, be clear that the condition
$\rD/\rmr\sim 1$ is \textit{not} guaranteed for any given point
of the $\CC$XCDM parameter space. In this paper we have not
addressed the computation of the probability distribution for the
two energy levels $\rD$ and $\rmr$ to be comparable at a
particular time of the cosmic evolution. In the $\CC$XCDM model
discussed here the complete solution, including the dynamics of
the energy density levels at a particular time, depends
deterministically on the set of values of the free parameters in
Eq. (\ref{parameterspace2}). If we would discretize the parametric
space, i.e. dividing it into small 3D elements, and then calculate
the dynamics for every volume element, it would be possible to
numerically construct the joint probability distribution for the
matter and (composite) dark energy density levels at some fixed
time. Since the volume of the parametric space is infinite, one
would need to introduce additional constraints in order to select
a finite volume of parametric space and make this calculation
useful. Although an explicit construction of this distribution
could indeed be performed after defining these constraints, our
presentation actually provided an alternative procedure which
should serve equally well our purposes; namely, we explicitly
computed that part of the total parameter space which results in
the wanted dynamics (that is to say, the ratio $r=\rD/\rmr$ being
bounded from above with the value of maximal $r$ within an order
of magnitude from the present one) and showing that it is
non-negligible. This is patent in Fig.\,\ref{fig2}, where the
"physical region" (viz. the one satisfying the aforementioned
requirements) is exhibited. What we have found is that the
dimensionality of the "physical region" is the same as the
dimensionality of the entire parametric space itself, in other
words: the ``physical region'' is of non-zero measure in the
space $(\Omega_{\Lambda}^0, \omega_X,\nu)$. Not only so, the
fractional volume (referred to the volume of the region where the
parameters stay within their ``natural'' values, see
Fig.\,\ref{fig2}) is sizeable, as shown by the various
projections onto the different planes displayed in  Fig.\,
\ref{fig3}-\ref{fig4}. This proves that no additional
fine-tunings are necessary to realize this scenario which is, in
that sense, natural within the $\CC$XCDM models. This fact,
combined with the nice features of the effective EOS of dark
energy (described in Section \ref{sect:eEOS}) at the redshifts
amenable to distant supernovae observations (and the compatibility
with the nucleosynthesis bounds), further reinforces the $\CC$XCDM
models as models with acceptable behavior of dark energy which at
the same time provide an acceptable explanation (or substantial
alleviation) of the coincidence problem.}

 \FIGURE[t]{
\mbox{\resizebox*{0.5\textwidth}{!}{\includegraphics{Ht.eps}}\ \
   \resizebox*{0.5\textwidth}{!}{\includegraphics{rt.eps}}}
\caption{\textbf{(a)}  The expansion rate of the $\CC$XCDM model
in terms of the cosmic time, $H=H(t)$, for the same parameter
choices as in Fig.\,\ref{fig10}a,b,c; \textbf{(b)}  The ratio
$r=\rD/\rmr$, in units of its present value $r_0$, as a function
of the cosmic time for the same scenarios as in (a). In both cases
the cosmic time is measured in Hubble time units $H_0^{-1}$. The
present time lies at $t_0=13.7\,Gyr$, i.e. $\sim 0.98\,H_0^{-1}$}

\label{fig11}}

\subsection{Transition from deceleration to acceleration} \label{sect:tredshift}

A precise measurement of the transition redshift $z_t$ from
deceleration into acceleration could also help to restrict the
parameters of a given model. The value of $z_t$ is obtained
theoretically by considering the change of sign of the
deceleration parameter from $q(z)>0$ to $q(z)<0$, where
\begin{equation}
\label{pa1} q(z)=-\frac{\ddot{a}\,a}{\dot{a}^2}
=-\frac{\ddot{a}}{a\,H^2(z;\nu)}=\frac12\,\left[(1+3\,\wm)\,\tilde{\Omega}_m
(z)+(1+3\,\we)\,\,\tilde{\Omega}_D (z) \right]\,.
\end{equation}
The cosmological functions $\tilde{\Omega}_D (z)$ and
$\tilde{\Omega}_m(z)$ are defined in (\ref{Omegas3}). Since the
the point $z_t$ is reached during the matter epoch, where
$\wm=0,\ \we<0$, we can define $z_t$ as the point that satisfies
$q(z_t)=0$, hence
\begin{equation}\label{zt}
\frac{\tilde{\Omega}_D
(z_t)}{\tilde{\Omega}_m(z_t)}=\frac{1}{3\,|\we|-1}\,.
\end{equation}
In the standard $\CC$CDM model the corresponding relation reads
${\tilde{\Omega}_D (z_t)}={\tilde{\Omega}_m(z_t)}/2$, which is
recovered from (\ref{zt}) after setting $\we=-1$ and identifying
$\OD$ with $\OLo$. Equivalently, the deceleration parameter can be
computed from the expansion rate $H(z)$ as follows,
\begin{equation}\label{pa2}
q(z)\,=\,-1+\frac12\,(1+z)\,\frac{1}{H^2(z)}\,\frac{dH^2(z)}{dz}\,,
\end{equation}
and $z_t$ satisfies $q(z_t)=0$. For the standard $\CC$CDM model
and using the favorite values $\OMo=0.3$ and $\OLo=0.7$ of the
cosmological parameters in the flat case, one obtains
\begin{equation}\label{zstar}
z_t=-1+\sqrt[3]{2\,\frac{\OL^0}{\OM^0}}\simeq 0.67\,.
\end{equation}
For the $\CC$XCDM model the expansion rate $H=H(z)$ is more
complicated and is of course sensitive to the basic set of
parameters (\ref{parameterspace2}) through the formula given in
the first expression in Eq.\,(\ref{solved1}). However, due to the
relatively wide experimental margin obtained for $z_t$ (cf. Riess
\textit{et al.} in \cite{Supernovae})
\FIGURE[t]{
\mbox{\resizebox*{0.7\textwidth}{!}{\includegraphics{ratcoma.eps}}}
\caption{Comparison of the transition redshift $z_t$
(deceleration-acceleration) for $\wX=-1.85$, $\OLo=0.75$) and
different values of $\nu$. } \label{fig13} }
\begin{equation}\label{qexp}
z_t^{\rm exp}=0.46\pm 0.13\,,
\end{equation}
it is not possible at the moment to firmly discriminate which
value of $\nu$ is preferred. Notice that the range (\ref{qexp})
excludes the standard model value (\ref{zstar}) by a bit more
than one standard deviation, but this is not very significant for
the moment. We expect that this situation will substantially
ameliorate in the future. In the meanwhile from Fig.\,\ref{fig13}
we test the sensitivity of the transition redshift $z_t$ to the
value of $\nu$ in the $\CC$XCDM. It seems to favor positive values
of $\nu$. It is interesting to note that the $z_t$ sensitivity to
$\nu$ is opposite in the $\CC$XCDM as compared to the pure running
$\CC$ model of Ref.\,\cite{RGTypeIa}. In the latter the values
$\nu<0$ are favored (see particularly Fig.\,4 of the second
reference in \,\cite{RGTypeIa}). This is an example of how the
observables can discriminate between two different realizations of
a cosmological RG model. Both models, the $\CC$XCDM and the
running $\CC$ model of \cite{RGTypeIa} are based on a similar RG
framework. However, in the latter matter is not self-conserved
(there is exchange of energy between matter and the running $\CC$)
whereas in the present model  matter is self-conserved and the
running $\CC$ can exchange energy only with the cosmon. Obviously,
it will be necessary a comparison of several high precision
observables (a task presumably feasible thanks to the next
generation of observational experiments) before we can
discriminate between the two kinds of RG cosmologies. Another
observable that should help to discriminate between models is the
EOS itself and the existence of a transition point $z^{*}$ where
the model might effectively change from quintessence to
phantom-like behavior or vice versa. For example, the RG model of
\cite{RGTypeIa} may have transitions from quintessence into
phantom, as was shown in Ref.\,\cite{SS1}, whereas the $\CC$XCDM
model can only have transitions of the opposite kind, or no
transition whatsoever if stopping is required (cf. Section
\ref{sect:crossing}). Let us also add that the RG model of
\cite{RGTypeIa} does not have any clue, in contradistinction to
the present $\CC$XCDM model, as to finding an explanation for the
cosmological coincidence problem because the ratio (\ref{ratio})
-- which in that model just reads $\rL(z)/\rM(z)$ -- is unbounded.

\FIGURE[t]{
\mbox{\resizebox*{0.5\textwidth}{!}{\includegraphics{figure8a.eps}}\
\ \ \
   \resizebox*{0.5\textwidth}{!}{\includegraphics{figure8b.eps}}}
\caption{Phase trajectories of the autonomous system
(\ref{autonomous1}) in the $(\OL,\OX)$ plane. Dashed lines show
the parts of the curves corresponding to positive redshifts (i.e.
our past), full lines indicate the parts between the present
moment and the stopping (if there is stopping) and dotted lines
denote the inaccessible part of the trajectory after the stopping.
\textbf{(a)} Example of scenario VII in
Section\,\ref{sect:solving2} (saddle point case) corresponding to
$\wX=-1.85$, $\nu=-\nu_0$. The eigenvector directions
(\ref{eigenvectors}) are also marked. The various trajectories
correspond to different choices of $\OLo$ indicated in the
figure; \textbf{(b)} Example of scenario IV with a node
determined by $(\nu,\OLo)$; in the figure $\OLo=-2$, $\nu=0.96$.
All trajectories are attracted towards the node (\ref{nunode}),
but they get stopped on the way to it. The point $\zeta=0$
corresponds to the present. The cosmic evolution is in the
$\zeta>0$ direction, see Eq.\,(\ref{zetav}). }\label{fig14} }
\FIGURE[t]{\\
\\

\mbox{\resizebox*{0.5\textwidth}{!}{\includegraphics{figure9a.eps}}\
\ \ \
\resizebox*{0.5\textwidth}{!}{\includegraphics{figure9b.eps}}}
\caption{Phase trajectories of the autonomous system
(\ref{autonomous1}) in the $(\OD,\Om)$ plane. Cases \textbf{(a)}
and \textbf{(b)} are respectively as in Figs.\,\ref{fig14}a,b but
here we plot the corresponding trajectories in the phase plane for
the total dark energy density versus matter density, i.e.
$(\OD,\OM)$. The meaning of the lines is as in Fig.\,\ref{fig14}.}
 \label{fig15}}

\subsection{Phase trajectories of the cosmological autonomous system} \label{sect:phasecurves}

Finally, in Fig.\,\ref{fig14} and \ref{fig15} we show some phase
trajectories of the autonomous system (\ref{autonomous1}) in two
different planes and for particular scenarios covering the two
types of situations (node and a saddle point). Due to the stopping
conditions discussed in Section\,\ref{sect:solving2}, in both of
these cases the trajectories become eventually truncated and
therefore either did not have a chance to reach the convergence
node, or alternatively fail to escape away to infinity,
respectively. In Figs.\,\ref{fig14}a,b we display several phase
trajectories in the $(\OL,\OX)$ plane covering these
possibilities. In Fig.\,\ref{fig14}a\, we have an example that
fits with scenario VII of Section\,\ref{sect:solving2}. The
trajectories that turn left and get stopped are those fulfilling
the condition (\ref{OLOM2}), whereas those turning right do not
fulfill it and hence do not stop.  In contrast, in
Fig.\,\ref{fig14}b\, we are dealing with scenario IV, where in the
absence of stopping the trajectories would be attracted to the
node (\ref{nunode}) for all values of $\wX$ in the quintessence
range ($\wX\gtrsim-1$). In this particular example, which
illustrates a situation where $\OLo$ and $\nu$ are both rather
large, the fixed point is well out of the figure. Even so the
trajectories show already a clear trend to get focused. The choice
of the parameters does satisfy the stopping condition
(\ref{cond1}) and so the trajectories get stopped well before
reaching the node. It can be clearly seen in the figure because
the ``stopped part'' of the trajectories is partially shown.

In Figs.\,\ref{fig15}a,b\, we represent respectively the same kind
of scenarios as in Fig.\,\ref{fig14}a,b\, except that we perform
the representation in the plane of total DE density versus matter
density. In Fig.\,\ref{fig15}a the trajectories $\OD$ satisfying
the condition (\ref{OLOM2}) are seen to get stopped when crossing
the straight line $\OD=-\OM$ corresponding to $r=-1$. The
trajectories that do not fulfill (\ref{OLOM2}) escape away into
unbounded values of $\OD$. Similarly, in Fig.\,\ref{fig15}b we
have a situation with the presence of a ``would be node'' which
focus all trajectories, but all of them get stopped because the
condition (\ref{cond1}) is fulfilled.

\section{Discussion and conclusions}
\label{sect:conclusions}

In this paper we have explored the possibilities offered to
cosmology by a non-trivial composite model of the dark energy
($\CC$XCDM). By composite model we mean that it contains at least
two entities of entirely different nature, i.e.  not just a
replica of a given one (e.g. two or more scalar fields). By
non-trivial we mean that these two components generally interact
with one another. From our point of view it is perfectly
conceivable that the observed DE is made out of a mixture of
various components (perhaps some scalar fields) stemming from
different sources of a fundamental theory, e.g. string theory.
However, in the most general case we have to contemplate the
possibility that one of the ingredients of the DE could be a
cosmological term, $\CC$, with its own dynamics and interactions
with the rest. In our model the DE components other than $\CC$
have been represented by $X$,  the ``cosmon''. We should stress
that the cosmon need not be a fundamental field, it could be an
effective representation of dynamical fields of various sorts, or
even the effective behavior of higher order curvature terms in the
effective action. In this paper we have presented a composite DE
model of this kind, which we have narrowed down to its simplest
non-trivial form. The dynamics of the variable cosmological term,
$\CC$, is linked to the renormalization group (RG) running in
QFT, and the effective entity $X$ interacts with $\CC$, but not
with matter, such that both the total DE and matter-radiation
densities are separately conserved. In fact, the cosmon (whose
ultimate nature is not necessary to specify in our discussion)
follows an evolution which is fixed by the RG dynamics of $\CC$
itself through the requirement of conservation of the total DE.
The particular modeling of the $\CC$ evolution that we have used\
follows the law $\delta\CC\sim H^2$, which is well-motivated in
the RG framework of Ref.\,\cite{JHEPCC1,RGTypeIa,SSS}. The
modeling could of course be different in other implementations of
the RG in Cosmology, or even without explicit reference to it. In
this sense the class of $\CC$XCDM models is surely more general.
However, the non-trivial implementation presented here furnishes
a ``proof of existence'' of promising scenarios of this kind, and
suggests that the idea of having a ``scaling $\CC$'' may
constitute a fruitful interface between QFT and
Cosmology\,\cite{JHEPCC1}. At the very least it gives a hint on
the immense capability of the $\CC$XCDM cosmologies to describe
the next generation of high precision data, even without making
explicit use of simple scalar field models of the DE with
particular forms of their potentials.

In the $\CC$XCDM context the $\CC$ term has a very different
status as compared to the $\CC$CDM model. While the main
observable quantity here is the total DE density $\rD=\rL+\rX$,
the $\CC$XCDM  model places $\CC$ in a fully dynamical context
together with the cosmon. Within this approach the puzzling
question, formulated within the standard $\CC$CDM model, of why
$\CC$ takes a particular value at present, and why this value is
so close to the matter energy density, does no longer apply. The
total DE is an evolving quantity through the entire history of
the Universe, and under appropriate conditions it traces closely
the matter density. This is so because in this model we can find
a large region of the parameter space where the expansion of the
Universe stops at some point in the future. In contradistinction
to the standard $\CC$CDM model, here the ratio $\rD/\rmr$ is not
divergent with the evolution, and it reaches a maximum at some
point in our future before the Universe stops and reverses its
motion. We may view the present state of our Universe in this
model as just being in transit to reach that maximum and the
subsequent stopping/turning point. At present that ratio is of
order $1$, but at the maximum it can be, say, ten times bigger.

Obviously, this framework provides some clue for understanding the
cosmological coincidence problem. From our point of view this
problem is not so much the issue of why $\rD/\rmr$ is of order one
at present, but rather of why $\rD/\rmr\rightarrow\infty$ in the
asymptotic future regime (as within the standard $\CC$CDM model).
If that ratio is bounded, as it is the case in the model under
consideration, then the Universe can be at any value below this
maximum, and this feature certainly improves the status of the
purported cosmological coincidence. Within the $\CC$XCDM model the
fact that $\rD/\rmr\sim 1$ can be looked upon as a reflex of the
present state of accelerated expansion of the Universe, which
started somewhere between $z=0.3-0.7$, cf. Eq.\, (\ref{qexp}).
This interpretation is not possible in the standard $\CC$CDM model
because once the situation $\rD/\rmr\sim 1$ is reached the ratio
further increases without end, and so it is hard to understand why
at present this ratio comes out to be of order one. It can only be
if we admit that we live in a very special moment, namely one
rather close to the start of the acceleration. Similar problems
are met by the quintessence and phantom models in the literature.
In the former case we cannot understand why the (slowly
ever-decreasing) quintessence field matches up the (fast
decreasing) matter density precisely right now in the context of
an eternal Universe, whereas in the latter we cannot figure out
why the (slowly ever-increasing) phantom field just catches up
with the matter density squarely around our time.

Quite in contrast, in the $\CC$XCDM model, with non-vanishing
$\nu$, the ratio $\rD/\rmr$ remained relatively small in the
remote past, being roughly proportional to $|\nu\,(1+\wX)|$. (We
can easily provide values of the parameters for which
$\rD/\rmr\sim 10\%$ at the nucleosynthesis era.)  Later on, it
started to be of order $1$ around the transition epoch
($z=0.3-0.7$) where the Universe changed from a decelerated
expansion into an accelerated one. From the early times up to the
present day the ratio increased slowly, and in the future it will
continue increasing, but it will never surpass a finite number,
e.g. of order $10$, until the Universe will stop and reverse its
motion (a few Hubble times after our time). The upshot is that
from the time of primordial nucleosynthesis till the remote point
in the future where the ratio $\rD/\rmr$ becomes maximum, it
changed only mildly, say from $0.1$ to $10$, being currently
around $1$, in contrast to the fantastic change predicted in the
$\CC$CDM case (viz. from $0$ to $\infty$) during its infinite
lifetime. Therefore, once the $\CC$XCDM Universe starts to
accelerate, the probability to live at a time where $r$ is of
order $1$ is essentially $1$. Obviously, this model can
significantly ameliorate the situation with the cosmological
coincidence problem, and it may actually serve as a prototype
example of how to reach an explanation for this ``coincidence''
through the addition of some $X$ to $\CC$CDM.

Furthermore, we have shown that the effective EOS behavior of the
$\CC$XCDM model can mimic to a high extent that of the standard
$\CC$CDM model (namely by providing an average value of the
effective EOS parameter $\we$ near $-1$) in essentially the
entire redshift interval where the SNAP experiment\,\cite{SNAP}
is going to search for distant supernovae in the near future.
Moreover, the possibility of interaction between the cosmon and
the cosmological term (which is gauged by the parameter $\nu$)
allows to modulate the effective EOS of the $\CC$XCDM model by
giving it a tilt into the quintessence or phantom regime
depending on the sign and size of $\nu$ and the value of the
barotropic index of the cosmon, $\wX$. With no interaction
($\nu=0$) between the two components of the DE, this would be
impossible. This shows that even though $\OL$ and $\OX$ may not
be individually observable, the interaction between $\CC$ and $X$
can be detected through the effective EOS features that it gives
rise to.

We wish to stand out another important signature of the model that
involves comparing the two main ranges of redshift observations.
Remarkably, while the behavior of the effective EOS of the
$\CC$XCDM model at small and intermediate redshifts (relevant to
the supernovae analysis) can be very close to the standard
$\CC$CDM model, it turns out that at the very high redshift range
(relevant for CMB considerations) the effective EOS of the DE is
similar to that of matter-radiation. In this range the DE
physical effects of the $\CC$XCDM model could be detected through
a ``renormalization'' correction of the cosmological mass
parameter $\Omo$ (which could be up to order $10\%$) as compared
with the value fitted from the supernovae data. It turns out that
most of the parametrizations used in the literature to
(simultaneously) fit the effective EOS of the DE at low and high
$z$ could be inadequate to describe the $\CC$XCDM model and other
promising RG cosmological models whose EOS behavior exhibits this
kind of features\,\cite{SS1}.

{We may wonder if an effective description of the composite dark
energy used in the $\CC$XCDM model can be performed in terms of
two or more interacting scalar fields. The problem essentially
boils down to finding an appropriate potential for the scalar
fields. Should the form of this potential be a very complicated
or a contrived one, the effective description in terms of scalar
fields would not be very appealing.  A simple example shows that
the scalar field description is not always attractive. Take the
special case of the $\CC$XCDM model in which $\nu=0$, i.e. the
dark energy components do not mutually interact; in particular,
we could think of one component being a negative $\CC$ and the
other being e.g. a quintessence component $X$. In this case an
effective description in terms of two non-interacting scalar
fields would not be especially useful since one of the fields
would have to be non-dynamical. If on the other hand they would
be taken as interactive, this would lead to an unnecessarily more
complicated scalar field description of the DE as compared to the
original $\CC$XCDM one, and therefore it could not be considered
as a better option. Furthermore, even if an effective scalar
field model is found which reproduces the global expansion of the
universe in the $\CC$XCDM model, local properties such as growth
of perturbations might be different in these models. As we
stressed before, the cosmon entity need not be a fundamental
field. In this sense the $\CC$XCDM model can be the effective
behavior of many possible constructions, which in all cases have
in common the existence of a conserved total DE density. This
warns us that we should be careful to avoid extracting too
precipitated conclusions on the ultimate nature (e.g. scalar
field) of the DE.}

Clearly, the $\CC$XCDM model can encompass very general behaviors
of the DE. Therefore, the fact that the standard $\CC$CDM with a
strictly constant $\CC$ seems to be in fairly good agreement with
the last round of high precision WMAP experiments\,\cite{WMAP3Y}
-- with the possibility that future SNAP data could independently
corroborate this result -- in no way can be considered a
definitive confirmation of the $\CC$CDM. As shown in this paper,
the $\CC$XCDM model could do in principle a very similar job, with
the bonus that it can offer a natural explanation for the
cosmological coincidence problem. It remains to study its
implications for the structure formation and CMB, but this is of
course a task beyond the scope of the present paper. For the
moment we have shown that the model can be perfectly tested by
performing measurements at all accessible ranges of cosmologically
significant redshifts (both low and high), and therefore we expect
that the next generation of high precision
experiments\,\cite{SNAP,PLANCK,DES} will allow to better pin down
the features of the effective EOS of the dark energy such that
the $\CC$XCDM parameters can be better determined.

We conclude with the following observation. The ``addition'' of a
dynamical cosmon $X$ to the standard $\CC$CDM model can be viewed
(or not) as a natural feature in DE model building.  But in
practice a cosmon-like entity (e.g. in quintessence regime) has
thrived alone through the literature quite successfully (as a full
substitute for $\CC$) for quite some time, following perhaps the
courageous -- and certainly longer -- tradition of $\CC$ itself.
On the other hand, one may ponder whether it is natural or not to
allow a dynamical $\CC$ term. However, the running of $\CC$, or of
any other parameter, may be viewed as a natural property of
Quantum Field Theory applied to Cosmology\,\cite{JHEPCC1}. At the
end of the day it seems that other cosmological problems could be
significantly smoothed out if the DE is permitted to be neither
pure $\CC$ nor pure $X$, but just some more or less democratic
combination of both, including some peaceful interaction between
them.

\vspace{0.3cm}

{\bf Acknowledgments:}  This work has been supported in part by
MECYT and FEDER under project 2004-04582-C02-01, and also by DURSI
Generalitat de Catalunya under project 2005SGR00564. JG was also
supported by project  BES-2005-7803. The work of HS is financed by
the Secretar\'\i a de Estado de Universidades e Investigaci\'on of
the Ministerio de Educaci\'on y Ciencia of Spain within the
program \textit{Ayudas para la mobilidad de profesores,
investigadores, doctores y tecn\'ologos extranjeros en Espa\~na}.
HS thanks the Dep. ECM of the Univ. of Barcelona for the
hospitality.


\newcommand{\JHEP}[3]{{\sl J. of High Energy Physics } {JHEP} {#1} (#2)  {#3}}
\newcommand{\NPB}[3]{{\sl Nucl. Phys. } {\bf B#1} (#2)  {#3}}
\newcommand{\NPPS}[3]{{\sl Nucl. Phys. Proc. Supp. } {\bf #1} (#2)  {#3}}
\newcommand{\PRD}[3]{{\sl Phys. Rev. } {\bf D#1} (#2)   {#3}}
\newcommand{\PLB}[3]{{\sl Phys. Lett. } {\bf #1B} (#2)  {#3}}
\newcommand{\EPJ}[3]{{\sl Eur. Phys. J } {\bf C#1} (#2)  {#3}}
\newcommand{\PR}[3]{{\sl Phys. Rep } {\bf #1} (#2)  {#3}}
\newcommand{\RMP}[3]{{\sl Rev. Mod. Phys. } {\bf #1} (#2)  {#3}}
\newcommand{\IJMP}[3]{{\sl Int. J. of Mod. Phys. } {\bf #1} (#2)  {#3}}
\newcommand{\PRL}[3]{{\sl Phys. Rev. Lett. } {\bf #1} (#2) {#3}}
\newcommand{\ZFP}[3]{{\sl Zeitsch. f. Physik } {\bf C#1} (#2)  {#3}}
\newcommand{\MPLA}[3]{{\sl Mod. Phys. Lett. } {\bf A#1} (#2) {#3}}
\newcommand{\CQG}[3]{{\sl Class. Quant. Grav. } {\bf #1} (#2) {#3}}
\newcommand{\JCAP}[3]{{\sl J. of Cosmology and Astrop. Phys. }{ JCAP} {\bf#1} (#2)  {#3}}
\newcommand{\APJ}[3]{{\sl Astrophys. J. } {\bf #1} (#2)  {#3}}
\newcommand{\AMJ}[3]{{\sl Astronom. J. } {\bf #1} (#2)  {#3}}
\newcommand{\APP}[3]{{\sl Astropart. Phys. } {\bf #1} (#2)  {#3}}
\newcommand{\AAP}[3]{{\sl Astron. Astrophys. } {\bf #1} (#2)  {#3}}
\newcommand{\MNRAS}[3]{{\sl Mon. Not.Roy. Astron. Soc.} {\bf #1} (#2)  {#3}}




\begin {thebibliography}{99}

\bibitem{Peebles84}  P.J.E. Peebles, \APJ {284}{1984}{439}.

\bibitem{Supernovae} A.G. Riess \textit{ et al.}, \AMJ {116} {1998} {1009};
 S. Perlmutter \textit{ et al.}, \APJ {517} {1999} {565};
R. A. Knop \textit{ et al.}, \APJ {598} {2003} {102}; A.G. Riess
\textit{ et al.} \APJ {607} {2004} {665}.

\bibitem{WMAP03} D.~N.~Spergel {\it et al.},
\textsl{Astrophys.\, J.\, Suppl.}\  {\bf 148} (2003) 175; See also
the WMAP Collaboration: \ {\tt http://map.gsfc.nasa.gov/}

\bibitem{LSS} M. Tegmark \textit{et al}, \PRD {69}{2004}{103501}.

\bibitem{WMAP3Y} \textit{WMAP three year results: implications for
cosmology}, D.N. Spergel \textit{et al.},
\texttt{astro-ph/0603449}.

\bibitem{TurnerWhite} M. S. Turner, M. J. White, \PRD
{56}{1997}{4439}, astro-ph/9701138.

\bibitem{SNAP}
See all the relevant information for SNAP in: http://snap.lbl.gov/

\bibitem{PLANCK} See the relevant information for PLANCK in:
http://www.rssd.esa.int/index.php?project=Planck

\bibitem{DES} The Dark Energy Survey Collaboration,
\texttt{astro-ph/0510346}; see also the DES website
http://www.darkenergysurvey.org/

\bibitem{Linde} J. Kratochvil, A. Linde, E. V. Linder, M. Shmakova, \JCAP
{0407} {2004} {001}, \texttt{astro-ph/0312183}.

\bibitem{Alam} U. Alam, V. Sahni, A.A. Starobinsky, \textit{JCAP}
{0406} (2004) {008}; U. Alam, V. Sahni, T.D. Saini, A.A.
Starobinsky, \MNRAS {354} {2004} {275}.

\bibitem{Hannestad} S. Hannestad, E. Mortsell,
\textit{JCAP} 0409 (2004) 001.

\bibitem{Jassal1} H.K. Jassal, J.S. Bagla,
T. Padmanabhan, {\em Mon. Not. Roy. Astron. Soc. Letters} {\bf
356} (2005) L11-L16, \texttt{astro-ph/0404378};  \PRD {72}{2005}
{103503}, \texttt{astro-ph/0506748}.

\bibitem{Jassal2}  K. M. Wilson, G.
Chen, B. Ratra, \texttt{astro-ph/0602321}; H.K. Jassal, J.S.
Bagla, T. Padmanabhan, \texttt{astro-ph/0601389}; G.B Zhao, J.Q.
Xia, B. Feng, X. Zhang\,,\MNRAS {367} {2006} {825-837},
\texttt{astro-ph/0603621}; S. Nesseris, L. Perivolaropoulos,
\PRD{72}{2005}{123519},\,\texttt{astro-ph/0511040}.

\bibitem{Phantom} R.R. Caldwell, \PLB {545} {2002} {23}, \texttt{astro-ph/9908168}; R.R.
Caldwell, M. Kamionkowski, N.N. Weinberg, Phys. Rev. Lett. 91
(2003) 071301, \texttt{astro-ph/0302506}; A. Melchiorri, L.
Mersini, C.J. Odman, M. Trodden, \PRD {68} {2003} {043509},
\texttt{astro-ph/0211522}; H. \v{S}tefan\v{c}i\'{c}, \PLB {586}
{2004} {5}, \texttt{astro-ph/0310904}; \EPJ {36} {2004} {523},
\texttt{astro-ph/0312484}; \PRD {71} {2005} {124036},
\texttt{astro-ph/0504518}; S. Nojiri, S.D. Odintsov, \PRD {70}
{2004} {103522}; S. Capozziello, S. Nojiri, S.D. Odintsov, \PLB
{632}{2006}{597}, \texttt{hep-th/0507182}; B. Feng, X. L. Wang, X.
M. Zhang, \PLB {607}{2005}{35}, \texttt{astro-ph/0404224}; B.
Feng, \texttt{astro-ph/0602156}.



\bibitem{weinRMP} S. Weinberg, \RMP {\bf 61} {1989}  {1}.

\bibitem{CCRev} See e.g.\,
V. Sahni, A. Starobinsky, \IJMP {A9} {2000} {373}; S.M. Carroll,
\textsl{Living Rev. Rel.} {\bf 4} (2001) 1; T. Padmanabhan, \PR
{380} {2003} {235}.

\bibitem{DEPaddy} T. Padmanabhan, \textit{Dark Energy: the Cosmological Challenge of
the Millennium}, {\emph Current Science},  {\bf 88} (2005) 1057,
\texttt{astro-ph/0411044}.

\bibitem{Copeland06} E.J. Copeland, M. Sami, S. Tsujikawa,
\textit{Dynamics of dark energy}, \texttt{hep-th/0603057}.

\bibitem{JHEPCC1}  I.L. Shapiro, J. Sol\`{a},
\JHEP {0202} {2002} {006},
 \texttt{hep-th/0012227}; \PLB {475} {2000} {236},
\texttt{hep-ph/9910462}.

\bibitem{SantFeliu} J. Sol\`{a}, \textsl{Nucl. Phys. Proc. Suppl.} \textbf{95}
(2001) 29, \texttt{hep-ph/0101134}


\bibitem{Xinesos} J.A. Gu, W-Y.P. Hwang \PRD {73} {2003} {023519}
\texttt{astro-ph/0106387}; R. Cardenas, T. Gonzalez, O. Martin, I.
Quiros \PRD {67} {2003} {083501}, \texttt{astro-ph/0206315}.


\bibitem{RGTypeIa}  I.L. Shapiro, J. Sol\`a, C. Espa\~na-Bonet,
P. Ruiz-Lapuente,  \PLB {574} {2003} {149},
\texttt{astro-ph/0303306};  \textit{JCAP} {0402} (2004) {006},
\texttt{hep-ph/0311171}; I.L. Shapiro, J. Sol\`a, \NPPS {127}
{2004} {71}, \texttt{hep-ph/0305279}; I. L. Shapiro, J. Sol\`a,
JHEP proc. AHEP2003/013, \texttt{astro-ph/0401015}.


\bibitem{Hellerman} S. Hellerman, N. Kaloper, L. Susskind,
\JHEP {0106}{2001}{003}, \texttt{hep-th/0104180}; W. Fischler, A.
Kashani-Poor, R. McNees, S. Paban, \JHEP {0107}{2001}{003},
\texttt{hep-th/0104181}; T. Banks, M. Dine  \JHEP {0110} {2001}
{012}, \texttt{hep-th/0106276}.

\bibitem{Dolgov} A.D. Dolgov, in: \textit{The very Early
Universe}, Ed. G. Gibbons, S.W. Hawking, S.T. Tiklos (Cambridge
U., 1982); F. Wilczek, \PR {104} {1984} {143}; T. Banks, \NPB
{249}{1985}{332}; L. Abbott, \textsl{Phys. Lett.} \textbf{B150}
(1985) 427; S. M. Barr, \PRD {36}{1987}{1691}; L.H. Ford,
\textsl{Phys. Rev.} \textbf{D 35} (1987) 2339; T.P. Singh and T.
Padmanabhan, \IJMP {A3}{1988}{1593}.

\bibitem{PSW}  R.D. Peccei, J. Sol\`{a}, C. Wetterich, \PLB {195} {1987}
{183}; C. Wetterich, \NPB {302} {1988} 668; J. Sol\`{a}, \PLB
{228} {1989} {317}; \IJMP {A5} {1990} {4225}.

\bibitem{Ratra} B. Ratra, P.J.E. Peebles, \PRD {37} {1988} {3406};  C. Wetterich,
\NPB {302} {1988} 668.

\bibitem{quintessence} R.R. Caldwell, R. Dave, P.J. Steinhardt, \PRL
{80} {1998} {1582}; For a review, see e.g. P.J.E. Peebles, B.
Ratra, \RMP {75} {2003} {559}, and the long list of references
therein.


\bibitem{braneworld} C. Deffayet, G.R. Dvali, G. Gabadadze, \PRD
{65}{2002}{044023}\,,\texttt{astro-ph/0105068}.

\bibitem{Chaplygin} A. Yu. Kamenshchik,
U. Moschella, V. Pasquier, \PLB {511}{2001}{265},
\texttt{gr-qc/0103004}; N. Bilic, G.B. Tupper, R.D. Viollier,
Phys. Lett. B 535 (2002) 17, astro-ph/0111325; J.C. Fabris,
S.V.B. Goncalves, P.E. de Souza, \textit{Gen. Rel. Grav.} {34}
(2002) 53, gr-qc/0103083; M.C. Bento, O. Bertolami, A.A. Sen,
\PRD {66} {2002} {043507}, \texttt{gr-qc/0202064}.

\bibitem{Barr} S. M. Barr, S. Ng, R. J. Scherrer, hep-ph/0601053.

\bibitem{cosm} I.L. Shapiro,  J. Sol\`{a}, \PLB {475} {2000} {236},
\texttt{hep-ph/9910462}.

\bibitem{Bohmer} C.G. Boehmer, T. Harko, \texttt{gr-qc/0602081}
(to appear in \textit{Mod. Phys. Lett. A}).


\bibitem{Peccei} R.D. Peccei, \PRD {71}{2005}{023527}, hep-ph/0411137; R. Fardon,
A.E. Nelson, N. Weiner, \JCAP {10} {2004} {005},
astro-ph/0309800; P. Gu, X. Wang, X. Zhang, \PRD{68}
{2003}{087301}, \texttt{hep-ph/0307148}; P.Q. Hung,
\texttt{hep-ph/0010126}.


\bibitem{HawkingEOS} S.W. Hawking, G.F.R. Ellis, \textit{The Large Scale Structure
of Space-Time} (Cambridge Univ. Press, 1973); S. M. Carroll, M.
Hoffman, M, Trodden, \PRD{68}{2003}{023509},
\texttt{astro-ph/0301273}.


\bibitem{LinderEff}  V. Sahni, T. D. Saini, A. A. Starobinsky, U. Alam,
\textit{JETP Lett.} {\bf 77} (2003) 201, \textit{Pisma Zh. Eksp.
Teor. Fiz.} {\bf 77} (2003) 249, \texttt{astro-ph/0201498};
\MNRAS {344}{2003}{1057}, \texttt{astro-ph/0303009}; J. A.
Frieman, D. Huterer, E. V. Linder, M. S. Turner, \PRD {67} {2003}
{083505}, \texttt{astro-ph/0208100}; E.V. Linder, \PRD
{70}{2004}{023511}, astro-ph/0402503;
 E.V. Linder, A. Jenkins, \MNRAS {346}{2003}{573},
 astro-ph/0305286.



\bibitem{SS1} J. Sol\`a, H. \v{S}tefan\v{c}i\'{c}, \PLB
{624}{2005}{147},\, \texttt{astro-ph/0505133}.

\bibitem{SS2}   J. Sol\`a, H. \v{S}tefan\v{c}i\'{c}, \MPLA {21} {2006}
{479}, \texttt{astro-ph/0507110}; {\sl J. Phys.} {\bf A39} (2006)
6753, \texttt{gr-qc/0601012};  J. Sol\`a, {\sl J. Phys. Conf.
Ser.} {\bf 39} (2006) 179, \texttt{gr-qc/0512030}.


\bibitem{Guberina06} B. Guberina, R. Horvat, H. Nikolic,
\texttt{astro-ph/0601598}.


\bibitem{Steinhardt} P.J. Steinhardt,  in: Proc. of the 250th Anniversary
Conference on Critical Problems in Physics, ed. V.L. Fitch, D.R.
Marlow, M.A.E. Dementi (Princeton Univ. Pr.,  Princeton, 1997).

\bibitem{InteractingQ} L. Amendola, \PRD {62}{2000}{043511},
\texttt{astro-ph/9908023}; L. Amendola, D. Tocchini-Valentini,
\PRD {64} {2001} {043509}, \texttt{astro-ph/0011243}; W. Zimdahl,
D. Pavon, \PLB {521}{2001}{133}, \texttt{astro-ph/0105479}; J.P.
Uzan, \PRD{59}{1999}{123510}, \texttt{gr-qc/9903004}.


\bibitem{OscillatingQ}  S. Dodelson, M. Kaplinghat, E. Stewart,
\PRL {85} {2000} {5276}, \texttt{astro-ph/0002360}; B. Feng, M.
Li, Y.S. Piao, X. Zhang, \PLB {634} {2006 }{101},
\texttt{astro-ph/0407432}; K. Griest, \PRD {66}{2002}{123501},
\texttt{astro-ph/0202052}; E. V. Linder, \APP {25}{2006}{167},
\texttt{astro-ph/0511415}; S. Nojiri, S. D. Odintsov,
\texttt{hep-th/0603062}.


\bibitem{Scherrer} R. J. Scherrer, \PRD {71} {2005} {063519}, \texttt{astro-ph/0410508}.

\bibitem{Book}  I.L. Buchbinder, S.D. Odintsov and I.L. Shapiro,
\textsl{Effective Action in Quantum Gravity}, IOP Publishing
(Bristol, 1992); N.D. Birrell and P.C.W. Davies, \textsl{Quantum
Fields in Curved Space}, Cambridge Univ. Press (Cambridge, 1982).



\bibitem{SSS} I.L. Shapiro, J. Sol\`a, H. \v{S}tefan\v{c}i\'{c},
\textit{JCAP} {0501} (2005) {012}\,, \texttt{hep-ph/0410095}.

\bibitem{Babic}
A. Babic, B. Guberina, R. Horvat, H. \v{S}tefan\v{c}i\'{c}, \PRD
{65} {2002} {085002}; B. Guberina, R. Horvat, H.
\v{S}tefan\v{c}i\'{c} \PRD {67} {2003} {083001}; A. Babic, B.
Guberina, R. Horvat, H. Stefancic, \PRD {71} {2005}
{124041},\texttt{astro-ph/0407572}; B. Guberina, R. Horvat, H.
\v{S}tefan\v{c}i\'{c}, \JCAP {05}{2005}{001},
\texttt{astro-ph/0503495}.



\bibitem{Reuter} A. Bonanno, M. Reuter, \PRD {65} {2002}
{043508};  E. Bentivegna, A. Bonanno, M. Reuter, \JCAP {01} {2004}
{001}.

\bibitem{Bauer} F. Bauer, \CQG {22} {2005} {3533}, \texttt{gr-qc/0501078}; F.
Bauer, \texttt{gr-qc/0512007}.


\bibitem{CCvariable}  M. Reuter, C. Wetterich,
\PLB {188} {1987} {38}; K. Freese, F. C. Adams, J. A. Frieman, E.
Mottola, \NPB{287}{1987}{797}; P.J.E. Peebles, B. Ratra, \APJ
{L17}{1988}{325}; J.C. Carvalho, J.A.S. Lima, I. Waga,
\textsl{Phys. Rev.} \textbf{D46} (1992) 2404;  O. Bertolami 1986,
\text{Nuovo Cim.} B93 (1986)36.

\bibitem{Overduin} See e.g. J.M. Overduin, F. I. Cooperstock, \PRD {58} {1998} {043506} and  R.G.
Vishwakarma, \CQG {18} {2001} {1159}.

\bibitem{Barreiro} T. Barreiro, E.J. Copeland, N.J. Nunes, \PRD
{61}{2000}{127301}

\bibitem{McInness} B. McInnes  \JHEP {0208}{2002}{029},
\texttt{hep-th/0112066}.

\bibitem{bulkvisco} J. Ren, X.H. Meng, \texttt{astro-ph/0602462}.

\bibitem{Ferreira97} P. G. Ferreira, M. Joyce, \PRD {58}{1998}{023503},
\texttt{astro-ph/9711102}.

\bibitem{Carneiro05} S. Carneiro, \IJMP {D14}{2005}{2201},
\texttt{gr-qc/0505121}.


\end{thebibliography}
\end{document}